*

# QoS-SLA-Aware Adaptive Genetic Algorithm for Multi-Request Offloading in Integrated Edge-Cloud Computing in Internet of Vehicles

Leila Ismail*, *Member, IEEE*, Huned Materwala, and Hossam S. Hassanein, *Fellow, IEEE*

***Abstract*— The Internet of Vehicles over Vehicular Ad-hoc Networks is an emerging technology enabling the development of smart city applications focused on improving traffic safety, traffic efficiency, and the overall driving experience. These applications have stringent requirements detailed in Service Level Agreement. Since vehicles have limited computational and storage capabilities, applications requests are offloaded onto an integrated edge-cloud computing system. Existing offloading solutions focus on optimizing the application's Quality of Service (QoS) in terms of execution time, and respecting a single SLA constraint. They do not consider the impact of overlapped multi-requests processing nor the vehicle's varying speed. This paper proposes a novel Artificial Intelligence QoS-SLA-aware adaptive genetic algorithm (QoS-SLA-AGA) to optimize the application's execution time for multi-request offloading in a heterogeneous edge-cloud computing system, which considers the impact of processing multi-requests overlapping and dynamic vehicle speed. The proposed genetic algorithm integrates an adaptive penalty function to assimilate the SLA constraints regarding latency, processing time, deadline, CPU, and memory requirements. Numerical experiments and analysis compare our QoS-SLA-AGA to random offloading, and baseline genetic-based approaches. Results show QoS-SLA-AGA executes the requests 1.22 times faster on average compared to the random offloading approach and with 59.9% fewer SLA violations. In contrast, the baseline genetic-based approach increases the requests' performance by 1.14 times, with 19.8% more SLA violations.**

***Index Terms*—Artificial Intelligence (AI), Cloud Computing, Computation Offloading, Constrained Optimization, Edge Computing, Genetic Algorithm (GA), Intelligent Transportation System, Internet of Things (IoT), Internet of Vehicles (IoV), Quality of Service (QoS), Service Level Agreement (SLA), Vehicular Ad-hoc Network (VANET)**

## I. Introduction

Internet of Vehicles (IoV) over Vehicular Ad-hoc Networks (VANETS) are self-organizing networks of vehicles equipped to exchange data between mobile vehicles and an infrastructure [1]. The vehicles act as smart nodes having sensing, computing, storage, and networking capabilities [2], [3]. Data exchange is realized using vehicle-to-vehicle (V2V), vehicle-to-roadside (V2R), vehicle-to-infrastructure (V2I), vehicle-to-cloud (V2C), and vehicle-to-pedestrian (V2P) communication. IoV provides mechanisms to develop applications for safe driving and efficient traffic management [4]; such applications include accident prevention, infotainment, real-time navigation, image processing and pattern recognition for autonomous driving. However, the vehicle's limited computation and storage capabilities hinder the deployment of these compute-intensive and time-critical applications.

Vehicular cloud computing (VCC) [5] has been developed to enable compute-intensive vehicles requests to be processed on remote cloud servers [6] to comply with the processing and resource requirements of Service Level Agreements (SLAs). However, latency requirements of communication-bound applications may be violated due to long-distance data transmission between vehicles and remote cloud servers.

Consequently, Vehicular Edge Computing (VEC) [7] pushed cloud services to the edge of the radio access network in closer proximity to the mobile vehicles, thus reducing communication delay. However, the VEC servers (deployed within Roadside Units (RSUs)) may violate the stringent deadline constraints of compute-intensive applications due to their limited computing capabilities. Consequently, it becomes necessary to develop a mechanism to offload vehicular requests onto an integrated edge-cloud computing system to comply with SLA requirements of latency for communication-bound applications (e.g., traffic alert and accident prevention) and processing for computation-bound applications (e.g., computer vision and multimedia), while optimizing applications' Quality of Service (QoS) [8].

Several works proposed computation offloading algorithms for an integrated edge-cloud computing system in the Internet of Things (IoT)/IoV networks. Most focus on optimizing the applications' QoS without considering SLA requirements [9]–[12]; or respecting SLA without QoS optimization [13]–[18]. Some works proposed QoS-SLA-aware offloading solutions

*This research was funded by the National Water and Energy Center of the United Arab Emirates University (Grant 31R215). *Corresponding author: Leila Ismail (e-mail: leila@uaeu.ac.ae).*

Leila Ismail and Huned Materwala are with the Intelligent Distributed Computing and Systems (INDUCE) Research Laboratory, Department of Computer Science and Software Engineering, College of Information Technology, United Arab Emirates University (UAEU), Al Ain, Abu Dhabi, 15551, United Arab Emirates (UAE), and with National Water and Energy Center, UAEU, Al Ain, Abu Dhabi, 15551, UAE.

Hossam S. Hassanein is with the School of Computing, Queen's University, Ontario, Canada.

[19]–[25]. To the best of our knowledge, no work has considered the impact of multi-requests overlapping in heterogeneous edge-cloud computing system servers. Thus, we propose a novel Artificial Intelligence QoS-SLA-aware adaptive genetic algorithm (QoS-SLA-AGA) for offloading vehicular requests. It aims to optimize the QoS by minimizing the total execution time of the vehicular requests while respecting SLAs in terms of latency (the total communication time of the request), processing time, deadline (the summation of communication and computation times), CPU, and memory requirements via an adaptive penalty function. The algorithm learns from selected search space solutions to find an improved one. In addition, the proposed offloading algorithm considers the impact of executing multiple requests in edge/cloud resources on application performance in IoV. The specific contributions of this work in the field of computational offloading in the IoV are the following:

- We formulate an optimization algorithm for multi-request offloading in a heterogeneous integrated edge-cloud computing system for IoV that minimizes the total execution time of the vehicular requests while respecting the requests' SLA requirements.
- We propose a novel QoS-SLA-AGA to solve the formulated constrained optimization problem via an adaptive penalty function.
- We perform a convergence analysis of the proposed GA to obtain the optimal values of GA's parameters.
- We compare the performance of the proposed algorithm with random offloading and baseline genetic-based approaches in terms of total execution time and SLA violations (SLAVs) with varying SLA requirements.

The remainder of this article is organized as follows. In section II, we discuss related work. Section III describes the system model of an integrated edge-cloud computing system for the IoV. We formulate the offloading optimization problem in Section IV. Section V explains our proposed QoS-SLA-AGA algorithm for offloading. Section VI presents our numerical experiments, comparative analysis, and performance results. Finally, we conclude and suggest future directions in Section VII.

## II. Related Work

We divide the current literature on offloading for an IoT-edge-cloud integrated computing system into three categories:

1) QoS-aware offloading that optimizes applications' QoS without considering SLA requirements [9]–[12],
2) SLA-aware offloading that respects applications' SLA constraints without enhancing the QoS [13]–[18], and
3) QoS-SLA-aware offloading that optimizes applications' QoS while respecting SLA constraints [19]–[25].

**QoS-aware offloading**, Pham et al. [9] proposed a game-theoretic approach to execute mobile device requests locally on the device or an edge server. The algorithm optimizes the weighted sum of the request's execution time and the device's energy consumption. However, the reply's communication time and the device's mobility are not considered. Xu et al. [10] proposed a decomposition-based evolutionary algorithm to offload vehicular requests on edge nodes such that the total execution time of the request is minimized and the resource utilization of the edge nodes is maximized. However, the mobility of the vehicle outside the communication range of the edge executing the request is not considered. Similarly, the deep learning and the Non-dominated Sorting GA (NSGA) III offloading algorithms in [11] and [12] respectively do not consider the mobility of the vehicles/devices when optimizing the requests' execution times and vehicles/devices energy. In addition, the algorithms in [9]–[12] do not consider the edge/cloud servers heterogeneity.

**SLA-aware offloading solutions**, the optimization algorithms in [13]–[17] schedule mobile devices' requests to either edge, edge/cloud, or mobile/edge to optimize requests' acceptance, edge/cloud profit, mobile devices and edges servers energy, requests acceptance along with edge operational cost, and network usage respectively. The algorithms used in these works are based on Lagrangian relaxation, simulated annealing, deep reinforcement learning, or heuristic. Wu et al. [18] proposed an offloading algorithm either locally on the IoT device or on the edge/cloud such that the total energy consumption of the IoT device is the minimum. These works consider a request's total execution time as a constraint. However, the algorithms [13]–[18] do not consider the devices' mobility and heterogeneity among the edge/cloud servers. Moreover, the communication time to deliver the request's reply is not considered in [15], [18].

**QoS-SLA-aware offloading solutions**, the algorithms in [19], [20] offload IoT devices' requests to edge/cloud using dynamic switching and fuzzy logic respectively. The requests' execution times are optimized in [19] and the execution time and edges' resource utilization in [20]. These algorithms consider the requests' execution times as constraints. Algorithms in [21]–[24] schedules vehicles' requests either on the vehicle, vehicle/edge, or vehicle/edge/cloud. The total execution time of the requests and the computational costs are optimized in [21], [22] using game theory, whereas the total execution time and load balancing on the edges are optimized using mixed-integer non-linear programming in [23]. Zhu et al. [24] minimize the weighted sum of maximum execution time and total quality loss using linear programming and particle swarm optimization. A request's execution time is considered as a constraint in [21]–[23], whereas the execution time, and the vehicle's available CPU and memory are considered as constraints in [24]. Peng et al. [25] proposed NSGA II and strength Pareto evolutionary algorithm to schedule mobile device's requests either locally or edge/cloud such that summation of the total execution time of the requests and the device's energy consumption is minimized while respecting the constraint on the request's execution time. However, the algorithms in [20], [21], [24], [25] do not consider the mobility of the IoT devices/vehicles, and [19], [22], [24], [25] do not consider communication time to deliver the reply of the request. The heterogeneity in the edge and cloud servers is not considered by [21], [22].

In summary, very few works have focused on QoS-SLA-aware offloading [19]–[25]. Only [23] considers memory requirements as SLA, and [22] considers the dynamic speed of the vehicle. To the best of our knowledge, no work considers the impact of multi-requests overlapping on the offloading decision and the dynamic speed of vehicles. In this paper, we propose a QoS-SLA-AGA for offloading in an integrated edge-cloud computing system for IoV that aims to improve the QoS by minimizing the total execution time of the applications' requests while respecting the SLAs in terms of latency, processing time, deadline, CPU, and memory requirements. Furthermore, our algorithm considers the impact of executing multiple requests in edge/cloud resources on application performance in IoV.

## III. System Model

Fig. 1 shows our integrated edge-cloud computing system model for vehicular networks that consists of three layers: 1) vehicles, 2) VEC, and 3) cloud computing. The first layer consists of $H$ vehicles moving with dynamic speed on a bi-directional road. Each vehicle $v_h$ $(h \in \mathcal{H})$ travels from a source to the destination location and has an application request $r_i$ $(i \in \mathcal{I})$ that should be executed. A request is represented as a tuple $r_i \triangleq \left(\psi_{r_i}, \sigma_{r_i}, \varphi_{r_i}, L_{r_i}^{max}, P_{r_i}^{max}, D_{r_i}^{max}, S_{v_h}(t), (x_{v_h,r_i}^{src}, y_{v_h,r_i}^{src}), (x_{v_h,r_i}^{des}, y_{v_h,r_i}^{des})\right)$. Requests in our system model are atomic and cannot be further divided into sub-requests. Consequently, each request can be executed on one at most one edge/cloud server. The requests vary in terms of computational requirement (i.e., length, CPU, and memory utilization values) and communication demand (i.e., data size).

The second layer (i.e., VEC) consists of $J$ RSUs placed alongside the road at equidistant. Each $RSU_j$ $(j \in \mathcal{J})$ has a coverage range of $D_{RSU}$ and is equipped with an edge server $e_j$ through a wired connection. The edge servers are heterogeneous in terms of processing and storage capabilities. A vehicle $v_h$ can communicate with an edge server $e_j$ only if it is under the communication range of $RSU_j$. We define a binary variable $\alpha_{v_h}^{e_j}(t) \in \{0,1\}$; such that $\alpha_{v_h}^{e_j}(t) = 1$ means that $v_h$ is in the range of $RSU_j$ and can communicate with $e_j$ and $\alpha_{v_h}^{e_j}(t) = 0$ otherwise. The third layer (i.e., cloud computing) consists of $K$ heterogeneous cloud servers such that the processing and storage capabilities of a cloud server $c_k$ $(k \in \mathcal{K})$ is higher compared to that of an edge server $e_j, \forall j \in \mathcal{J}$, i.e., $\mu_{e_j} \ll \mu_{c_k}$ and $\theta_{e_j} \ll \theta_{c_k}$.

Each edge server in our model receives a set of requests from the communicating vehicles. The server makes the offloading decision for each request $r_i$, i.e., to execute the request locally on the edge $e_j$ or to offload it to a cloud server $c_k$ for execution such that the total execution time of all the requests is at the minimum while maintaining each request's latency, processing time, deadline, CPU, and memory SLA constraints. A binary variable $\beta_{r_i}^{s_z} \in \{0,1\}, s_z \in \{e_j, c_k\}$ is defined. $\beta_{r_i}^{s_z} = 0$ if $r_i$ is executed locally on the edge server $e_j$ and $\beta_{r_i}^{s_z} = 1$ otherwise. For each offloaded request, the edge server sends the request and information about the cloud server $c_k$ to the cloud manager. The cloud manager then schedules the request to $c_k$. Since both communication and computation are critical for making the offloading decision, we next introduce the communication and computation models in detail. Table I lists the notations used in this paper and their definitions.

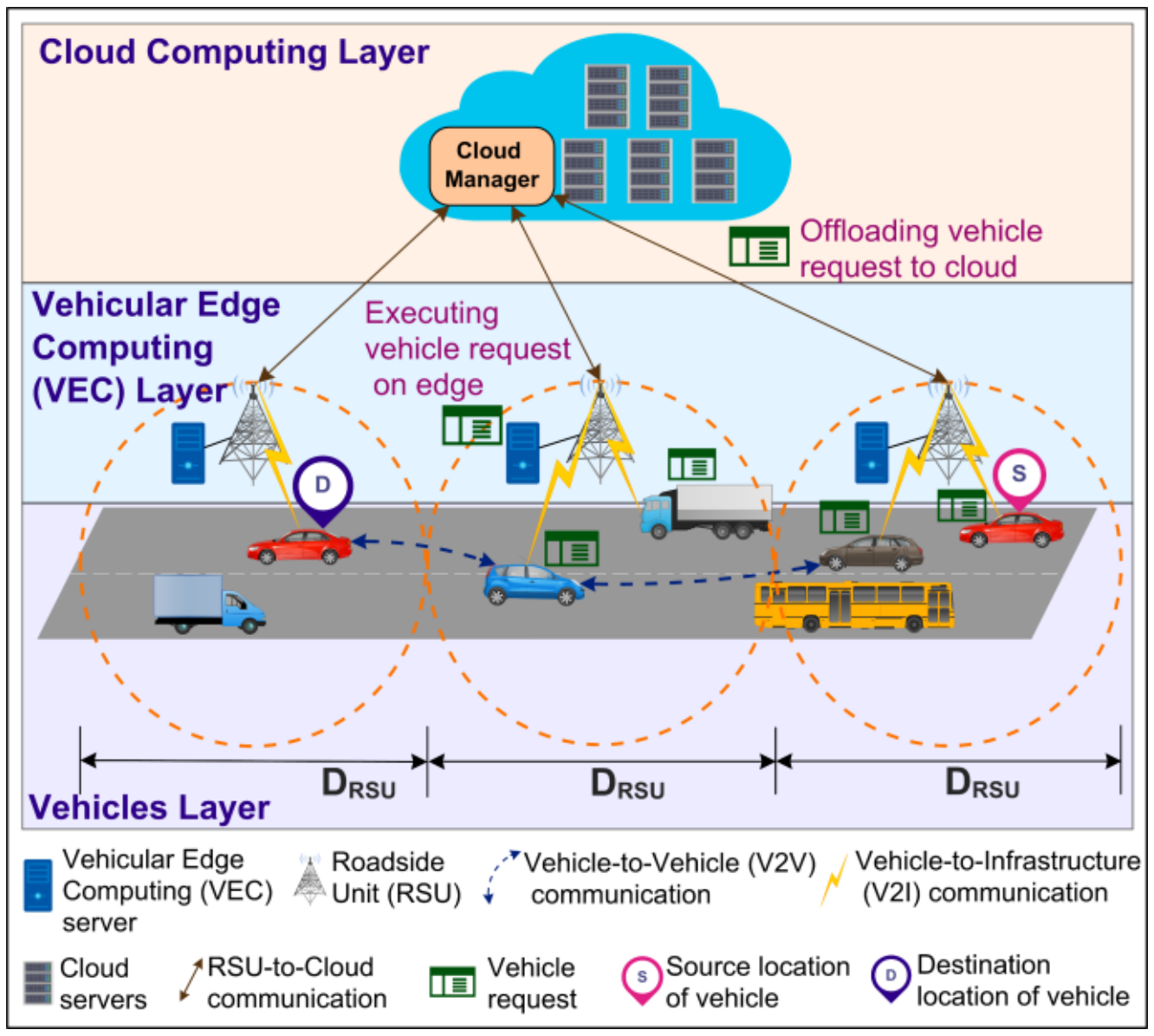

**Fig. 1.** Integrated edge-cloud computing system model for the Internet of Vehicles.

**Table I:** Notations and Definitions.

| Notation | Definition |
|---|---|
| $h, H, \mathcal{H}, v_h$ | vehicle index, number of vehicles, set of vehicles, $h^{th}$ vehicle |
| $i, I, \mathcal{I}, r_i, rep_i$ | request index, number of requests, set of requests, $i^{th}$ request, the reply of $r_i$ |
| $j, J, \mathcal{J}, e_j, RSU_j$ | edge server/RSU index, number of edge servers/RSUs, set of edge servers/RSUs, $j^{th}$ edge server, $j^{th}$ RSU |
| $k, K, \mathcal{K}, c_k$ | cloud server index, number of cloud servers, set of cloud servers, $k^{th}$ cloud server |
| $z, \mathcal{Z}, s_z$ | server (edge/cloud) index ($z = \{j, k\}$), set of edge and cloud servers ($\mathcal{Z} = \{\mathcal{J}\} \cup \{\mathcal{K}\}$), $z^{th}$ server $\left(s_z = \{e_j, c_k\}\right)$ |
| $\psi_{r_i}$ | length of $r_i$ in Million Instructions (MI) |
| $\sigma_{r_i}$ | size of $r_i$ in kilobytes (KB) |
| $\varphi_{r_i}$ | CPU utilization of $r_i$ |
| $L_{r_i}^{max}$ | maximum tolerable latency for $r_i$ |
| $P_{r_i}^{max}$ | maximum tolerable processing time for $r_i$ |
| $D_{r_i}^{max}$ | maximum tolerable deadline for $r_i$ |
| $S_{v_h}(t)$ | speed of $v_h$ at time $t$ |
| $(x_{v_h,r_i}^{src}, y_{v_h,r_i}^{src})$ | source location (longitude, latitude) of $v_h$ while submitting $r_i$ |
| $(x_{v_h,r_i}^{des}, y_{v_h,r_i}^{des})$ | destination location (longitude, latitude) of $v_h$ that submitted $r_i$ |
| $D_{RSU}$ | the coverage area of each RSU |
| $\mu_{s_z}$ | Processing speed of $s_z$ in Million Instructions per Second (MIPS) |
| $\theta_{s_z}$ | available memory of $s_z$ in KB |
| $\alpha_{v_h}^{e_j}(t)$ | whether $v_h$ can communicate with $e_j$ |
| $\beta_{r_i}^{s_z}$ | whether $r_i$ is executed locally on $e_j$ or offloaded to $c_k$ |
| $T_{r_i(s_z)}^{com}$ | total communication time of $r_i$ when executed on $s_z$ |
| $T_{r_i(x,y)}^{com}$ | communication time to transfer $r_i$ from $x$ to $y$, where $x \in \{v_h, e_j\}$ and $y \in \{e_j, c_k\}$ |
| $T_{rep_i(x,y)}^{com}$ | communication time to transfer $rep_i$ from $x$ to $y$, where $x \in \{e_j, c_k, e_{j+y}\}$ and $y \in \{v_h, c_k, e_{j+y}\}$ |
| $T^{proc}_{r_i(s_z)}$ | processing time of $r_i$ when executed on $s_z$ |
| $T^{I/O}_{r_i(s_z)}$ | I/O time of $r_i$ when executed on $s_z$ |
| $T_{r_i(s_z)}$ | total execution time of $r_i$ when executed on $s_z$ |
| $\omega_{x,y}$ | bandwidth between $x$ and $y$ in gigabits per seconds (Gbps), where $x, y \in \{v_h, e_j, c_k, e_{j+y}\}$ |
| $d_{v_h,e_j}$ | distance traveled by $v_h$ in the communication range of $e_j$ before submitting a request |
| $d^{r_i-rep_i}_{v_h}(t)$ | total distance traveled by $v_h$ after submitting $r_i$ and before receiving $rep_i$ |
| $\tilde{d}^{r_i-rep_i}_{v_h}(t)$ | distance traveled by $v_h$ outside the range $e_j$ of after submitting $r_i$ and before receiving $rep_i$ |
| $x^{left}_{e_j}$ | longitude of the left coordinate of $e_j$ such that $x^{left}_{e_j} = x^{src}_{v_h,r_i}$ |
| $x^{right}_{e_j}$ | longitude of the right coordinate of $e_j$ such that $x^{right}_{e_j} = x^{src}_{v_h,r_i}$ |
| $\tau^{m}_{r_i(s_z)}$ | processing time of $r_i$ when overlapped with other requests on $s_z$ |
| $\tau^{a}_{r_i(s_z)}$ | processing time of $r_i$ after other overlapping requests on $s_z$ has finished execution |
| $n_{r_i(s_z)}$ | number of requests executing along with $r_i$ on $s_z$ |
| $\bar{n}_{r_i(s_z)}$ | number of requests that were overlapping $r_i$ on $s_z$ and has completed execution before $r_i$ |
| $\chi_{r_i(s_z)}$ | number of times request data for $r_i$ is swapped between disk and memory on $s_z$ |
| $\xi_{s_z}$ | time required to transfer data between disk and memory in $s_z$ |
| $\rho_{r_i(s_z)}$ | the ratio of memory required by $r_i$ and the available memory on $s_z$ |
| $q, Q$ | offloading solution index, set of offloading solutions in a generation of genetic algorithm |
| $\ddot{F}_q, \tilde{\ddot{F}}_q$ | non-penalized fitness score of $q$, normalized non-penalized fitness score of $q$ |
| $P^{lat}_q$ | latency violation of $q$ |
| $P^{proc}_q$ | processing time violation of $q$ |
| $P^{deadline}_q$ | deadline violation of $q$ |
| $P^{cpu}_q$ | CPU violation of $q$ |
| $P^{mem}_q$ | memory violation of $q$ |
| $\tilde{P}_q$ | normalized SLA violations of $q$ |
| $\bar{F}_q, F_q$ | adaptive penalized fitness score of $q$, final fitness score of $q$ |
| $n_f$ | number of feasible offloading solutions in a generation |
| $\gamma$ | the ratio of feasible and total offloading solutions in a generation |
| $P_{size}$ | population size, i.e., number of offloading solutions, in a generation |
| $Cumm(q)$ | cumulative fitness probability of $q$ |
| $Prob(q)$ | fitness probability of $q$ |
| $\lambda_c, \lambda_m$ | crossover rate, mutation rate |
| $n_{mut}$ | number of requests for which the server allocation is mutated |

A. Communication Model

The communication time in our model consists of the time required to transmit the request data from $v_h$ to a scheduled server and the time required to transmit the reply back to $v_h$. Depending on the offloading decision (i.e., edge or cloud execution) and the speed of $v_h$, there are three scenarios as shown in Fig. 2. Fig. 2 (a) shows scenario (1) where $r_i$ is executed locally on $e_j$ and $v_h$ is connected to $e_j$ when a reply on $r_i$ is sent to $v_h$. In this scenario, the communication time involves the request transfer time from $v_h$ to $e_j$ and the reply transfer time from $e_j$ to $v_h$. Fig. 2 (b) shows scenario (2) where $r_i$ is executed on $e_j$ and $v_h$ is in the range of some edge server $e_{j+y}$ when a reply is sent to $v_h$. The communication time in this scenario includes the request transfer time from $v_h$ to $e_j$, the reply transfer time from $e_j$ to cloud server $c_k$, the reply transfer time from $c_k$ to $e_{j+y}$, and the reply transfer time from $e_{j+y}$ to $v_h$ [26]. The data transmission between $e_j$ and $e_{j+y}$ is achieved via the cloud instead of multi-hop RSU transmission. This is because multi-hop RSU transmission is performed at a low rate which increases the response time of the request [27]. Scenario (3) shown in Fig. 2 (c) indicates that $r_i$ is offloaded to the cloud and executed on a cloud server $c_k$. In this scenario, the communication time includes the request transfer time from $v_h$ to $e_j$, the request transfer time from $e_j$ to $c_k$, the reply transfer time from $c_k$ to $e_{j+y}$, and the reply transfer time from $e_{j+y}$ to $v_h$. In Fig. 2 (c), the vehicle moves out of $e_j$'s range before receiving a reply. Consequently, $c_k$ sends the reply to an edge server $e_{j+y}$. In case the vehicle is in the range of $e_j$, the reply will be transmitted from $c_k$ to $e_j$. Based on these scenarios, the total communication time for request $r_i$ when executed on a server $s_z \mid s_z \in \{e_j, c_k\}$ can be computed as stated in Equation 1.

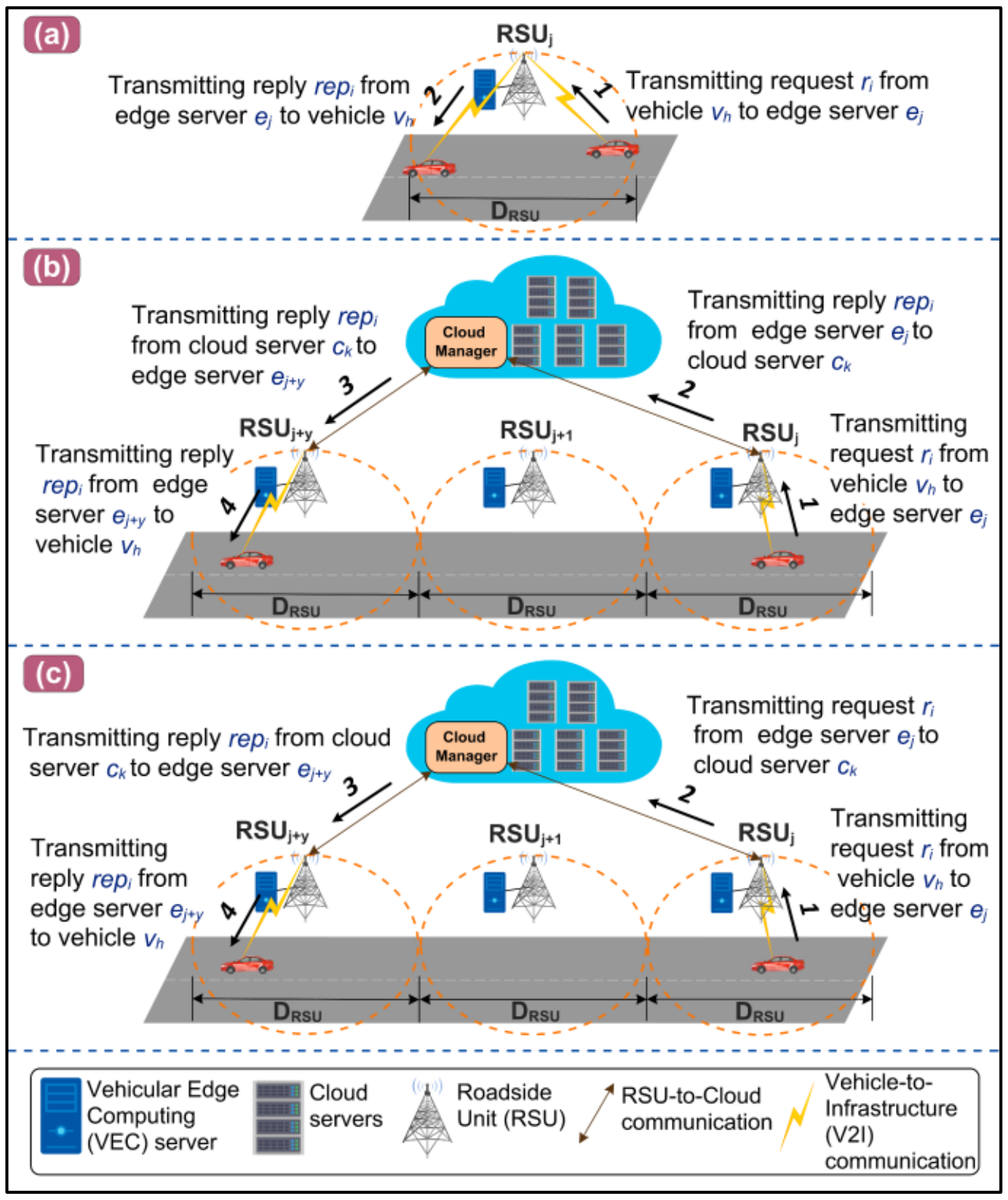

**Fig. 2.** Calculation of communication time based on offloading decision and vehicle speed. (a) vehicle request is executed on an edge server and vehicle is in the edge server's range when a reply on the request is received, (b) vehicle request is executed on an edge server and vehicle moves out of the edge server's range before receiving a reply on the request, and (c) vehicle request is executed on a cloud server.

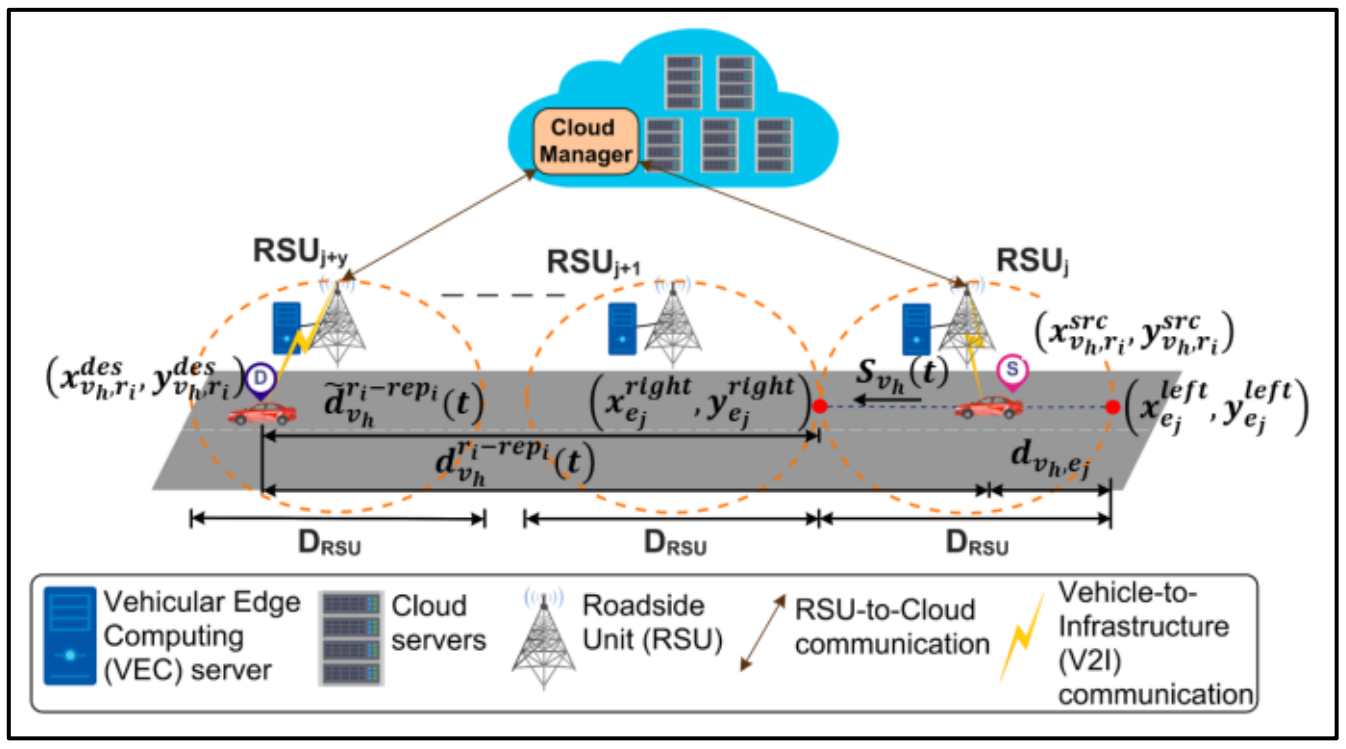

**Fig. 3.** Calculation of vehicle's position to send a reply on the request.

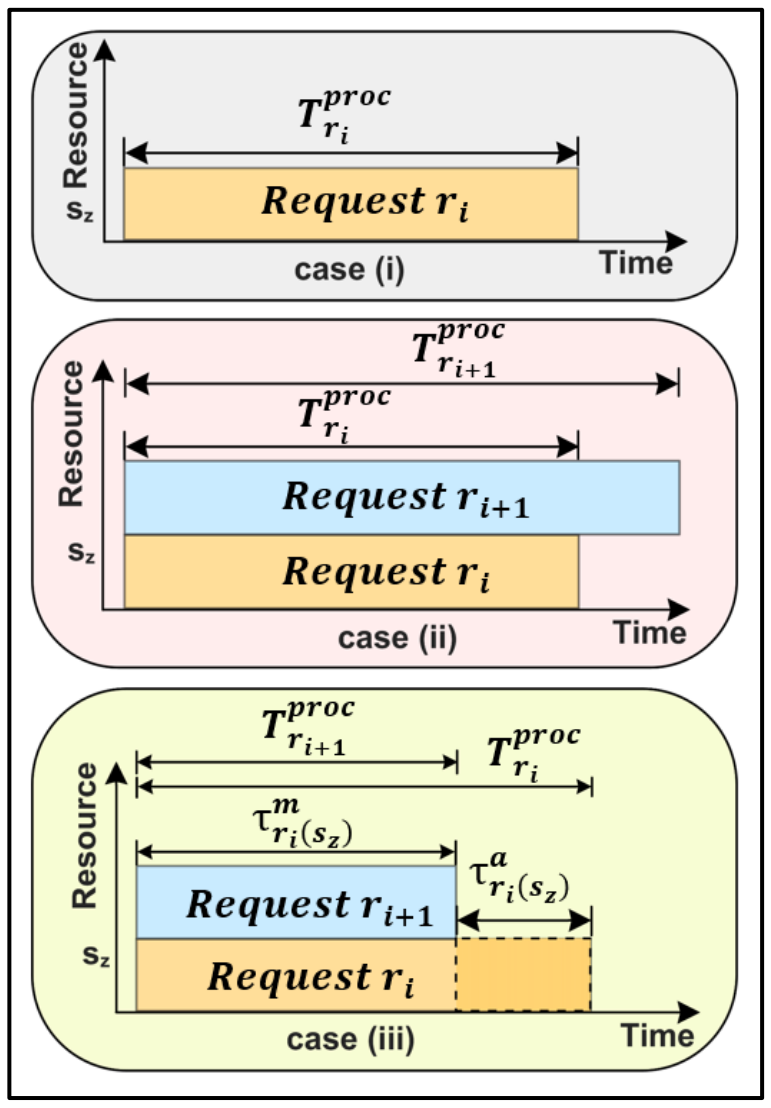

**Fig. 4.** Calculation of processing time in a multi-request scenario.

$$T^{com}_{r_i(s_z)} = \begin{cases} T^{com}_{r_i(v_h,e_j)} + T^{com}_{rep_i(e_j,v_h)}; & (s_z = e_j)\ and\ \left(\alpha^{e_j}_{v_h} = 1\right)\ while\ receiving\ the\ reply \\ T^{com}_{r_i(v_h,e_j)} + T^{com}_{rep_i(,e_j,c_k)} + T^{com}_{rep_i(c_k,e_{j+y})} + T^{com}_{rep_i(e_{j+y},v_h)}; & (s_z = e_j)\ and\ \left(\alpha^{e_j}_{v_h} = 0\right)\ while\ receiving\ the\ reply \\ T^{com}_{r_i(v_h,e_j)} + T^{com}_{r_i(,e_j,c_k)} + T^{com}_{rep_i(c_k,e_{j+y})} + T^{com}_{rep_i(e_{j+y},v_h)}; & s_z = c_k \end{cases} \tag{1}$$

The communication times represented in Equation 1 can be computed using Equations 2 – 7.

$$T^{com}_{r_i(v_h,e_j)} = \frac{\sigma_{r_i}}{\omega_{v_h,e_j}} \tag{2}$$

$$T^{com}_{rep_i(e_j,v_h)} = \frac{\sigma_{rep_i}}{\omega_{e_j,v_h}} \tag{3}$$

$$T^{com}_{rep_i(e_j,c_k)} = \frac{\sigma_{rep_i}}{\omega_{e_j,c_k}} \tag{4}$$

$$T^{com}_{rep_i(c_k,e_{j+y})} = \frac{\sigma_{rep_i}}{\omega_{c_k,e_{j+y}}} \tag{5}$$

$$T^{com}_{rep_i(e_{j+y},v_h)} = \frac{\sigma_{rep_i}}{\omega_{e_{j+y},v_h}} \tag{6}$$

$$T^{com}_{r_i(e_j,c_k)} = \frac{\sigma_{r_i}}{\omega_{e_j,c_k}} \tag{7}$$

In scenarios (2) and (3), the cloud manager should determine the RSU on the path between $v_h$'s source and destination under whose communication range $v_h$ will be when a reply is sent, i.e., $RSU_{j+y}$. The path between the source and destination can be computed by the cloud offline using an extended A* algorithm [28]. The algorithm determines an optimal path between source and destination in a way that reduces fuel consumption and travel time. The selection of the A* algorithm is based on its performance compared to the shortest path algorithm. Each vehicle in our system model will transmit its speed to the communicating edge server. The edge server will further transmit the speed to the cloud manager. To determine $RSU_{j+y}$, the manager will compute $d_{v_h,e_j}$, $d^{r_i-rep_i}_{v_h}(t)$, and $\tilde{d}^{r_i-rep_i}_{v_h}(t)$ as stated in Equations 8 - 10 (Fig. 3). The $y^{th}$ RSU after $RSU_j$, i.e., $RSU_{j+y}$ can be then determined using Equation 11. The value of $y$ will be updated in real-time based on the speed of $v_h$.

$$d_{v_h,e_j} = \begin{cases} x^{src}_{v_h,r_i} - x^{left}_{e_j}; & x^{src}_{v_h,r_i} < x^{des}_{v_h,r_i} \\ x^{right}_{e_j} - x^{src}_{v_h,r_i}; & x^{src}_{v_h,r_i} > x^{des}_{v_h,r_i} \end{cases} \tag{8}$$

$$d^{r_i-rep_i}_{v_h}(t) = s_{v_h}(t) \times T_{r_i(s_z)} \tag{9}$$

$$\tilde{d}^{r_i-rep_i}_{v_h}(t) = \begin{cases} d^{r_i-rep_i}_{v_h}(t) - \left(D_{RSU} - d_{v_h,e_j}\right); & d_{v_h,e_j} + d^{r_i-rep_i}_{v_h}(t) > D_{RSU} \\ 0; & otherwise \end{cases} \tag{10}$$

$$y(t) = \left\lceil \frac{\tilde{d}^{r_i-rep_i}_{v_h}(t)}{D_{RSU}} \right\rceil \tag{11}$$

*B. Computation Model*

The computation time in our system model includes the processing time to execute the request on an edge/cloud server and the I/O time required for the request data transfer between memory and disk of an edge/cloud server. The processing time and I/O time are explained below in detail.

*B.1. Processing Model*

The processing time of a request in our system model depends on whether the request is executed alone or with other requests on an edge/cloud server. Consequently, there exist three cases as shown in Fig. 4. The cases are as follows.

- Case (i): request $r_i$ is executed alone on an edge/cloud server $s_z$.
- Case (ii): execution of requests $r_i$ and $r_{i+1}$ overlap on server $s_z$ and $r_i$ completes execution before $r_{i+1}$. In this case, the processing speed of $s_z$ is divided among requests $r_i$ and $r_{i+1}$ [29].
- Case (iii): execution of requests $r_i$ and $r_{i+1}$ overlap on server $s_z$ and $r_i$ completes execution after $r_{i+1}$. In this case, the processing time of $r_i$ includes time $\tau^{m}_{r_i(s_z)}$ when execution of $r_i$ and $r_{i+1}$ overlaps and time $\tau^{a}_{r_i(s_z)}$ when $r_i$ executes alone after $r_{i+1}$ has finished execution.

The processing time for the three cases can be computed as stated in Equation 12.

$$T^{proc}_{r_i(s_z)} = \begin{cases} \left(\dfrac{\psi_{r_i}}{\mu_{s_z}}\right); & case\ (i) \\ \left(\dfrac{\psi_{r_i} \times n_{r_i(s_z)}}{\mu_{s_z}}\right); & case\ (ii) \\ \tau^{m}_{r_i(s_z)} + \tau^{a}_{r_i(s_z)}; & case\ (iii) \end{cases} \tag{12}$$

where $\tau^{m}_{r_i(s_z)}$ , $\tau^{a}_{r_i(s_z)}$, and $\bar{n}_{s_z}$ can be calculated using Equations 13 – 15.

$$\tau^{m}_{r_i(s_z)} = \min_{\forall r_p \in s_z} \left(T^{proc}_{r_p(s_z)}\right), p \neq i \tag{13}$$

$$\tau^{a}_{r_i(s_z)} = \left(\frac{\psi_{r_i} - \left(\dfrac{\tau^{m}_{r_i(s_z)} \times \mu_{s_z}}{n_{r_i(s_z)}}\right)}{\mu_{s_z}}\right) \times \left(n_{r_i(s_z)} - \bar{n}_{r_i(s_z)}\right) \tag{14}$$

$$\bar{n}_{r_i(s_z)} = \underset{r_i, r_p \in s_z}{\#} \left(T_{r_p(s_z)} < T_{r_i(s_z)}\right), i \neq p \tag{15}$$

*B.2. I/O Model*

The I/O time for a request $r_i$ on server $s_z$ in our system model refers to the time it takes to transfer data between disk and memory in case the memory requirement of $r_i$ is more than the available memory of $s_z$. The I/O time of $r_i$ when executed on $s_z$ for the three cases discussed before can be computed using Equation (16).

$$T^{I/O}_{r_i(s_z)} = \begin{cases} \left(\chi_{r_i(s_z)} \times \xi_{s_z}\right); & cases\ (i)\ and\ (ii) \\ \xi_{s_z} + \dfrac{\sigma_{r_i} \times \xi_{s_z} \times \bar{n}_{r_i(s_z)}}{\theta_{s_z}}; & case\ (iii) \end{cases} \tag{16}$$

where $\chi_{r_i(s_z)}$ is calculated using Equations 17 and 18 as follows.

$$\chi_{r_i(s_z)} = \begin{cases} \rho_{r_i(s_z)} - 1; & \rho_{r_i(s_z)} > 1 \\ 0; & \rho_{r_i(s_z)} \leq 1 \end{cases} \tag{17}$$

$$\rho_{r_i(s_z)} = \begin{cases} \left\lceil \dfrac{\sigma_{r_i}}{\theta_{s_z}} \right\rceil; & case\ (i) \\ \left\lceil \dfrac{\sigma_{r_i} \times n_{r_i(s_z)}}{\theta_{s_z}} \right\rceil; & cases\ (ii)\ and\ (iii) \end{cases} \tag{18}$$

## IV. Problem Formulation

We formulate a computation offloading optimization problem in an integrated edge-cloud computing system for IoV. The objective of the optimization problem is to minimize the total execution time of all the requests in the system under specified latency, processing time, deadline, CPU, and memory requirements constraints for each request. The total execution time of a request $r_i$ is computed as the summation of the request's communication, processing, and I/O times as stated in Equation 19. To this end, the corresponding optimization problem can be formulated as stated in Equation 20.

$$T_{r_i(s_z)} = T^{com}_{r_i(s_z)} + T^{proc}_{r_i(s_z)} + T^{I/O}_{r_i(s_z)}, \forall i \in \mathcal{I}, s_z \epsilon \{e_j, c_k\} \tag{19}$$

**Problem:**

$$minimize \sum_{\forall i \in \mathcal{I}} T_{r_i(s_z)}, \ \ s_z \in \{e_j, c_k\} \tag{20}$$

***s.t.***

C1: $\forall i \in \mathcal{I}\ T^{com}_{r_i(s_z)} \leq L^{max}_{r_i}, s_z \in \{e_j, c_k\}$

C2: $\forall i \in \mathcal{I}\ T^{proc}_{r_i(s_z)} \leq P^{max}_{r_i}, s_z \in \{e_j, c_k\}$

C3: $\forall i \in \mathcal{I}\ T_{r_i(s_z)} \leq D^{max}_{r_i}, s_z \in \{e_j, c_k\}$

C4: $\sum_{\forall r_i \in s_z} \varphi_{r_i} \leq \varphi^{max}_{s_z}, s_z \in \{e_j, c_k\}\ and\ \varphi_{r_i} \geq 1$

C5: $\sum_{\forall r_i \in s_z} \sigma_{r_i} \leq \theta_{s_z}, z \in \mathcal{Z}$

C6: $\alpha^{e_j}_{v_h}(t) \in \{0,1\}, \forall j \in \mathcal{J}, \forall h \in \mathcal{H}$

C7: $\sum_{j \in \mathcal{J}} \alpha^{e_j}_{v_h}(t) = 1, \forall h \in \mathcal{H}$

C8: $\sum_{\forall z \in Z} \beta^{s_z}_{r_i} = 1, \forall i \in \mathcal{I}$

Here $\sum_{\forall i \in \mathcal{I}} T_{r_i(s_z)}$ is the total execution time of all the requests in the system. The constraints in the above optimization problem are as follows:

- C1 ensures that the communication time of each request does not exceed the maximum tolerable latency requirement of that request.
- C2 guarantees that the execution time of each request is less than the request's permissible maximum processing time requirement.
- C3 ensures that the total execution time of each request is below the maximum tolerable deadline for that request.
- C4 ensures that the total CPU utilization of all the requests executing on a server should not exceed the server's CPU utilization threshold. This is to ensure that the server is not overloaded as overloading may degrade the requests' performances.
- C5 is the constraint on the memory resource, i.e., the

memory requirements of a request should be less than the server's available memory. This is to reduce the amount of data transfer between disk and memory.

- C6 and C7 denote that each vehicle can only communicate with one edge server at a given time.
- C8 ensures that each request is executed at most by only one edge/cloud server.

The optimization problem (Equation 20) for computation offloading is NP-hard where the search space and time to obtain the optimal solution increase exponentially with increasing requests and servers (edge and cloud). Hence, we propose an adaptive genetic algorithm to obtain the optimal solution in polynomial time.

## V. Proposed QoS-SLA-Aware Adaptive Genetic Offloading Algorithm

The proposed offloading aims to minimize the total execution time of the vehicles' requests while respecting each request's SLA requirements. In this paper, we propose a QoS-SLA-AGA to obtain the solution of the NP-hard offloading algorithm. GA [30] is based on the theory of natural evolution where a subset of near-optimal offloading solutions from one generation is used to obtain the offspring solution for the next generation. At each generation, the algorithm converges towards the global optima. In the context of our optimization problem, global optima can be defined as an offloading solution that yields the minimum requests execution time while respecting each request's SLA requirements. An offloading solution, referred to as chromosome in genetic algorithm terminology, consists of server allocation for each request. Each request-server allocation is known as a gene. The number of offloading solutions in each generation represents the population size ($P_{size}$) and remains constant throughout the generations. In the following, we explain the steps involved in our proposed algorithms.

### *A. Initialization of Offloading Solutions*

In this step, a set $Q$ of initial offloading solutions for the first generation is randomly developed to begin the exploration process in the search space. The number of solutions generated is equal to $P_{size}$. The value of $P_{size}$ should be carefully selected as it impacts the convergence of the algorithm. A small value improves the computational performance of the algorithm, however, may restrict the search space leading to local optima instead of global. On the other hand, a large value allows the algorithm to explore a larger search space that might lead to global optima. However, this increases the computational time.

### *B. Evaluation of the Offloading Solutions*

In this step, each offloading solution is evaluated, in terms of fitness, to determine how close it is to the optimal offloading solution. The closer a solution is to the optimal solution, the higher is its fitness. Consequently, based on our optimization objective function stated in Equation 20, an offloading solution having the least total execution time with no SLA violations will have the highest fitness value. To incorporate the SLA constraints in fitness computation, we implement an adaptive penalty function [31] that reduces the fitness value of an offloading solution that violates SLA requirements, referred to as an infeasible solution. To evaluate the fitness with an adaptive penalty, we first compute the non-penalized fitness value and the constraints violations for each solution as stated in Equations 21 – 26.

$$\ddot{F}_q = \sum_{\forall i \in \mathcal{I}} T_{r_i(s_z)}, \quad s_z \in \{e_j, c_k\}, \forall q \in Q \tag{21}$$

$$P_q^{lat} = \sum_{\forall T_{r_i(s_z)}^{com} > L_{r_i}^{max}} T_{r_i(s_z)}^{com} - L_{r_i}^{max} \tag{22}$$

$$P_q^{proc} = \sum_{\forall T_{r_i(s_z)}^{proc} > P_{r_i}^{max}} T_{r_i(s_z)}^{proc} - P_{r_i}^{max} \tag{23}$$

$$P_q^{deadline} = \sum_{\forall T_{r_i(s_z)} > D_{r_i}^{max}} T_{r_i(s_z)} - D_{r_i}^{max} \tag{24}$$

$$P_q^{cpu} = \sum_{\forall r_i \in s_z} I_{\left(\varphi_{r_i} > \varphi_{s_z}^{max}\right)}, \forall z \in \mathcal{Z} \tag{25}$$

$$P_q^{mem} = \sum_{\forall r_i \in s_z} I_{\left(\sigma_{r_i} > \theta_{s_z}\right)}, \forall z \in \mathcal{Z} \tag{26}$$

The non-penalized fitness and the constraints violations are normalized as using Equations 27 and 28 respectively. The adaptive penalized fitness is then calculated as stated in Equation 29.

$$\tilde{\ddot{F}}_q = \frac{\ddot{F}_q - \min_{\forall q \in Q}\left(\ddot{F}_q\right)}{\max_{\forall q \in Q}\left(\ddot{F}_q\right) - \min_{\forall q \in Q}\left(\ddot{F}_q\right)} \tag{27}$$

$$\tilde{P}_q = \frac{1}{5}\left(\frac{P_q^{lat}}{\max_{\forall q \in Q}\left(P_q^{lat}\right)} + \frac{P_q^{proc}}{\max_{\forall q \in Q}\left(P_q^{proc}\right)} + \frac{P_q^{deadline}}{\max_{\forall q \in Q}\left(P_q^{deadline}\right)} + \frac{P_q^{cpu}}{\max_{\forall q \in Q}\left(P_q^{cpu}\right)} + \frac{P_q^{mem}}{\max_{\forall q \in Q}\left(P_q^{mem}\right)}\right) \tag{28}$$

$$\bar{F}_q = \begin{cases} \tilde{P}_q; & n_f = 0 \\ \tilde{\ddot{F}}_q; & \tilde{P}_q = 0 \\ \sqrt{\left(\tilde{\ddot{F}}_q\right)^2 + \left(\tilde{P}_q\right)^2} + \left[(1-\gamma)\tilde{P}_q + (\gamma)\tilde{\ddot{F}}_q\right]; & otherwise \end{cases} \tag{29}$$

The final fitness score for each solution is then computed by taking the reciprocal of adaptive penalized fitness as stated in Equation 30. This is to assign the highest fitness value to the offloading solution having the least execution time and QoS violations.

$$F_q = \frac{1}{\bar{F}_q + 1} \tag{30}$$

### *C. Selection of Offloading Solutions to Reproduce Solutions for Next Generation*

In this step, offloading solutions from the population are selected based on their fitness value to reproduce offspring offloading solutions for the next generation. In this paper, we use the fitness proportionate Roulette Wheel Selection (RWS) [32] method that constructs a roulette wheel based on the cumulative fitness probabilities of the offloading solutions. The fittest a solution is, the larger the area occupied by that solution on the roulette wheel. The cumulative probability for each offloading solution can be computed using Equation 31. Offloading solutions are then selected based on the position of randomly generated numbers on the roulette wheel.

$$Cumu(q) = \sum_{l=1}^{q} Prob(l) \tag{31}$$

Where $Prob(q)$ represents the fitness probability of offloading solution $q$ $(q \in \mathcal{Q})$ and can be computed using Equation 32.

$$Prob(q) = \frac{F_q}{\sum_{\forall q \in \acute{O}} F_q} \tag{32}$$

*D. Crossover to develop next Generation Offloading Solutions*

In this step, the selected fit offloading solutions are used to produce offspring solutions by swapping the request-server allocations for two offloading solutions, known as parent solutions. Crossover operation produces fitter offspring offloading solutions from fit parent solutions leading to convergence of algorithm towards the optimal solution. The number of parent solutions selected for crossover depends on the crossover rate $\lambda_c$. In this paper, we use a single point crossover where a cutoff point for crossover is generated randomly and all the server allocations for the requests after the cutoff point from the parents are swapped resulting in two offspring solutions. The parent solutions in the generation are then replaced by the two fittest solutions among the parent and offspring solutions.

*E. Mutation to Diversify the Offloading Solutions in a Generation*

In this step, the offloading decisions for some requests in the population are changed to diversify the offloading solutions and larger the search space. Without mutation, the algorithm may converge prematurely, i.e., on the local optima, as the search space would be restricted around the non-optimal fit solutions in the population. The number of requests for which the offloading decisions are changed depends on the mutation rate parameter and can be calculated using Equation 3.

$$n_{mut} = I \times P_{size} \times \lambda_m \tag{33}$$

*F. Termination of the Algorithm*

In this step, the algorithm is terminated if the maximum number of user-defined generations is reached, or the optimal offloading solution is obtained. The evaluation, selection, crossover, and mutation operations are iterated until termination.

## VI. Performance Evaluation

We analyze the impact of different genetic algorithm parameters on the convergence of our algorithm and compare its performance with baseline approaches in terms of total execution time, and the number of requests violating SLA constraints.

*A. Experimental Environment*

We created a heterogeneous integrated edge-cloud computing system for IoV. Ten edge servers and 20 cloud servers were simulated using the different types of edge and cloud servers listed in Table II. Servers 1 and 2 originate from the Intelligent Distributed Computing and Systems (INDUCE) Research Laboratory, College of Information Technology, United Arab Emirates University. The specifications of the remaining servers, 3 – 6 are taken from the SPEC Power benchmark such that they belong to the same family of servers in our laboratory but with different capabilities. We implemented the network using MATLAB 2020a.

In our simulated network, we use the Vehicle-Crowd Interaction (VCI) – DUT dataset [33] for vehicles' positioning. In particular, we used the x_est and y_est columns of the dataset for the source and destination locations of vehicles in our experiments. Regarding the characteristics of the vehicular requests, we used three different ITS applications; facial recognition for autonomous driving, augmented reality, and infotainment [20], [34]. The network and application characteristics used in the experiments are listed in Table III. Table IV shows the values used for convergence analysis of the proposed QoS-SLA-AGA.

**Table II:** Specifications of the Servers used in the Experiments.

| Server | Location | Specification | Memory |
|---|---|---|---|
| 1 | Edge | AMD Opteron 252, 2.59 GHz, 2-Cores | 2GB |
| 2 | Cloud | Intel Xeon, 2.80 GHz, 2-Cores | 4GB |
| 3 | Edge | AMD Opteron 6276, 2.30 GHz, 16-Cores [35] | 32GB |
| 4 | Cloud | Intel Xeon E3-1204L v5, 2.10 GHz, -Cores [36] | 16GB |
| 5 | Cloud | Intel Xeon E-2176G, 3.7 GHz, 6-Cores [37] | 16GB |
| 6 | Cloud | AMD Opteron 6238, 2.60 GHz, 12-Core [38] | 64GB |

**Table III:** Network and Application Characteristics used in the Experiments.

| Parameter | Value(s) |
|---|---|
| Number of requests ($\mathcal{I}$) | 20, 25, 30, 35, 40, 45, 50 |
| Vehicle – RSU bandwidth (Gbps) $\left(\omega_{v_h,e_j}, \forall z \in \mathcal{Z}, \forall h \in \mathcal{H}\right)$ | 1 |
| RSU – cloud bandwidth (Gbps) $\left(\omega_{e_j,c_k}, \forall z \in \mathcal{Z}, \forall h \in \mathcal{H}\right)$ | U(1,2) |
| Time required for data swapping operation (seconds) $\left(\xi_{s_z}, \forall s_z \in \{e_j, c_k\}\right)$ | 0.05 |
| Server's CPU utilization threshold $\left(\varphi_{s_z}^{max}, \forall s_z \in \{e_j, c_k\}\right)$ | 90 |
| Requests' CPU utilization (%) $\left(\varphi_{r_i}, \forall i \in \mathcal{I}\right)$ | N(20, 5) |
| Requests' length (Million Instructions) $(\psi_{r_i}, \forall i \in \mathcal{I})$ | 9000 -15000 [20], [34] |
| Requests' size (KB) $\left(\sigma_{r_i}, \forall i \in \mathcal{I}\right)$ | 1000 – 5000 [20] |
| Requests' latency requirements (seconds) $\left(L_{r_i}^{max}, \forall i \in \mathcal{I}\right)$ | 0.1, 0.3, 0.5, 0.7, 0.9, 1.1 |
| Requests' processing time requirements (seconds) $\left(P_{r_i}^{max}, \forall i \in \mathcal{I}\right)$ | 0.9, 1.1, 1.3, 1.5, 1.7, 1.9 |
| Requests' deadline requirements (seconds) $\left(D_{r_i}^{max}, \forall i \in \mathcal{I}\right)$ | 1, 1.2, 1.4, 1.6, 1.8, 2 |

U denotes uniform distribution; N denotes a standard normal distribution

**Table IV:** Genetic Algorithm Parameters used for Convergence Analysis.

| Parameter | Value(s) |
|---|---|
| Crossover rate ($\lambda_c$) | 0.50, 0.55, 0.60, 0.65, 0.70, 0.75, 0.80, 0.85, 0.90, 0.95 |
| Mutation rate ($\lambda_m$) | 0.01, 0.02, 0.03, 0.04, 0.05, 0.06, 0.07, 0.08, 0.09, 0.1 |
| Population size ($P_{size}$) | $2 \times requests$, $4 \times requests$, $6 \times requests$, $8\times requests$, $10 \times requests$ |
| Termination condition | 1000 iterations (generations) |

### *B. Experiments*

This section explains the experiments performed for convergence analysis of QoS-SLA-AGA and compares its performance with baseline approaches in terms of total execution time and the number of requests violating SLA constraints.

To analyze the convergence of the proposed algorithm we executed the algorithm with different values of $\lambda_c$, $\lambda_m$, and $P_{size}$ (listed in Table IV).

1. We first run the algorithm with varying values of $\lambda_c$ and keep the values of $\lambda_m$ and $P_{size}$ constant at 0.01 and $2 \times requests$ respectively.
2. We then select the value of $\lambda_c$ that has the fastest convergence as the optimal crossover rate.
3. Next, we run the proposed algorithm with varying values of $\lambda_m$ and the values of $\lambda_c$ and $P_size$ constant at optimal crossover rate and $2 \times requests$ respectively.
4. We select the value of $\lambda_m$ resulting in the fastest convergence as the optimal mutation rate.
5. Lastly, we vary the values of $P_{size}$ with $\lambda_c$ and $\lambda_m$ constant at their optimal values.
6. We then select the value with the least convergence time as the optimal value for $P_{size}$.

We evaluate the offloading performance of the proposed algorithm with varying values of latency, processing time, deadline requirements, and number of requests (Table III). We vary one of those four parameters while keeping the remaining three constants at their corresponding minimum values. The value for deadline requirement is varied while varying the latency and processing time requirements, because the deadline is the summation of latency and processing times. For each run, we use the optimal values of $\lambda_c$, $\lambda_m$, and $P_{size}$. We measure the algorithm's performance in terms of the total execution time of the requests and the number of requests violating SLA requirements. For latency, processing time, and deadline violations, we calculate the number of requests for which the value of communication time, processing time, and total execution time is greater than the requests' requirements. To determine the number of requests violating CPU requirements, we consider the number of requests scheduled on a server where the total CPU utilization of all requests scheduled on that server exceeds the server's CPU utilization threshold.

To demonstrate the performance of the proposed QoS-SLA-AGA, we compare it with the following baseline approaches:

1) **QoS-Aware Genetic Algorithm (QoS-GA):** An offloading scheme using a genetic algorithm whose objective is to minimize the total execution time of all the requests without considering the SLA constraints.
2) **Random Offloading:** An offloading scheme where each request is randomly scheduled at the edge or cloud server without considering the QoS and SLA.

We repeat the experiments for QoS-GA and random offloading with varying latency, processing time, deadline requirements, and the number of requests.

### *C. Experimental Results Analysis*

This section presents the analysis of the results obtained from our experiments. In particular, we analyze the results on the convergence of our proposed algorithm and compare our algorithm with baseline approaches.

#### *C.1. Convergence Analysis*

Fig. 5 shows the convergence of QoS-SLA-AGA in terms of penalized fitness score distribution over iterations with varying values of $\lambda_c$. As shown, all the distributions over iterations are left-skewed, for all values of $\lambda_c$. However, $\lambda_c = 0.95$, has the highest mean fitness score of 0.997. Consequently, value $\lambda_c = 0.95$ converges the algorithm in the least amount of time. This is because a higher crossover rate diversifies the population by selecting more offloading solutions to perform the crossover operation. On the other hand, the offloading solutions are not as diverse when the crossover rate is low. Fig. 6 shows the distribution of total execution time of the requests over iterations with varying values of $\lambda_c$. As shown in the figure, $\lambda_c = 0.85$ which results in the fastest convergence with the shortest total execution time. Although, $\lambda_c = 0.85$ optimizes the total execution time, it does not provide the best convergence of the algorithm in terms of fitness (Fig. 5). This is because the fitness score in Fig. 5 considers both the optimization of total execution time and SLA violations, whereas Fig. 6 only considers the optimization of total execution time. Consequently, $\lambda_c = 0.85$ optimizes the time but violates SLA constraints. On the other hand, $\lambda_c = 0.95$ converges to an offloading solution with minimum execution time while respecting the SLA constraints. Consequently, we use $\lambda_c = 0.95$ in the remaining experiments.

Fig. 7 shows fitness convergence of QoS-SLA-AGA with varying values of $\lambda_m$. As depicted in the figure, only $\lambda_m = 0.01$ converges the algorithm to an optimal fitness value of 1. This is because a higher mutation rate hinders the convergence as fitter offloading solutions are lost. Fig. 8 depicts the distribution of total execution time over iterations with varying values of $\lambda_m$. As shown in the figure, $\lambda_m = 0.01$ converges the algorithm to the minimum execution time. No other values of $\lambda_m$ aids in the convergence of the algorithm. Consequently, we use $\lambda_m = 0.01$ in the offloading experiments.

Fig. 9 shows the fitness convergence of QoS-SLA-AGA with a varying value of $P_{size}$. As shown, the algorithm converges to the optimal fitness score of 1 with the highest mean when the population size is set to twice the number of requests. As shown in Fig. 10, the total execution time of the requests converges to the minimum quickly when $P_{size} =$ $2 \times requests$. Consequently, we use $P_{size} = 2 \times requests$ in the experiments. Table V shows the optimal genetic parameters used to evaluate the performance of the proposed QoS-SLA-AGA and baseline QoS-GA algorithms.

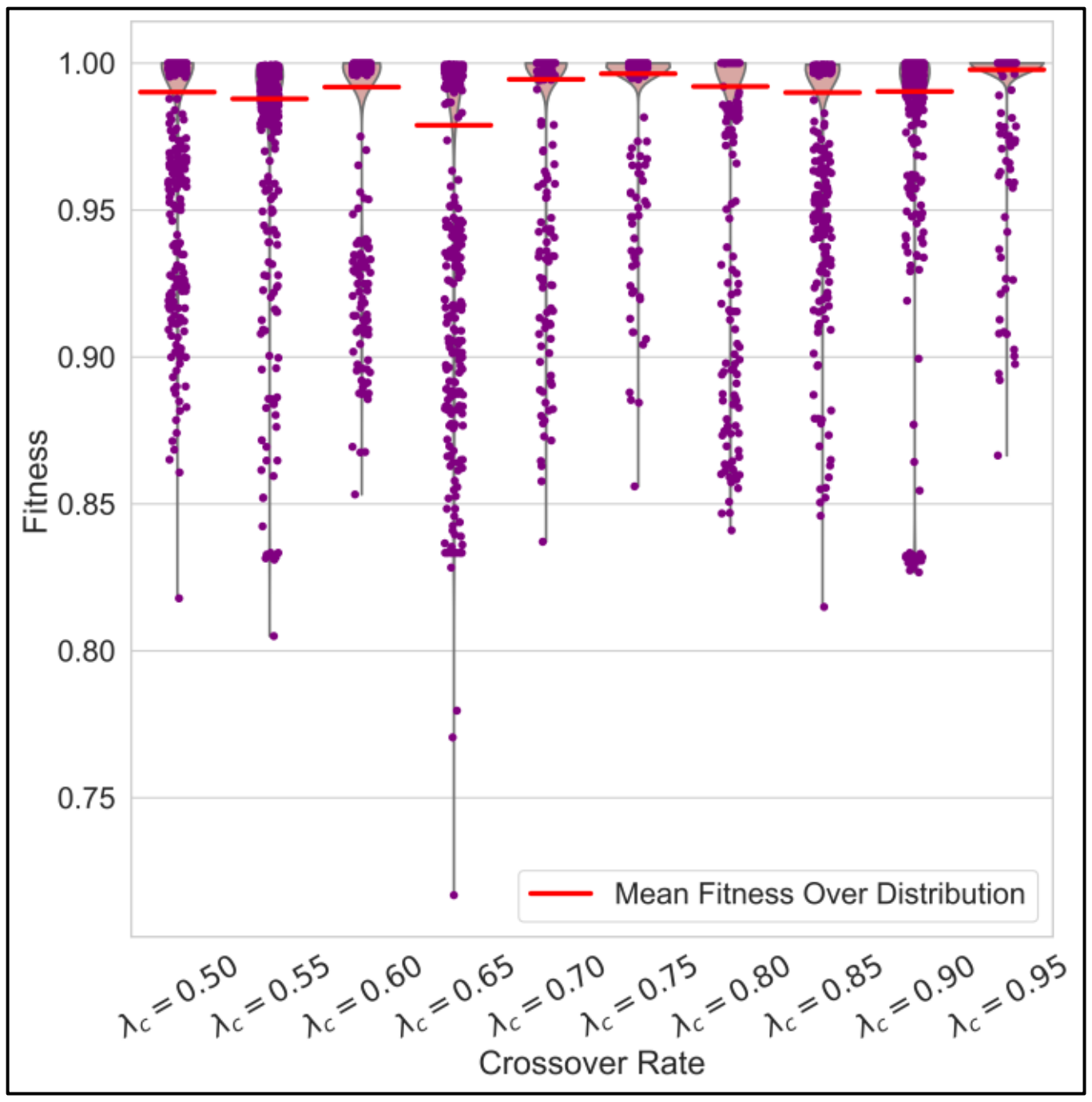

**Fig. 5.** Fitness score distribution over iterations for the requests' offloading solution using QoS-SLA-AGA versus crossover rate $\lambda_c$.

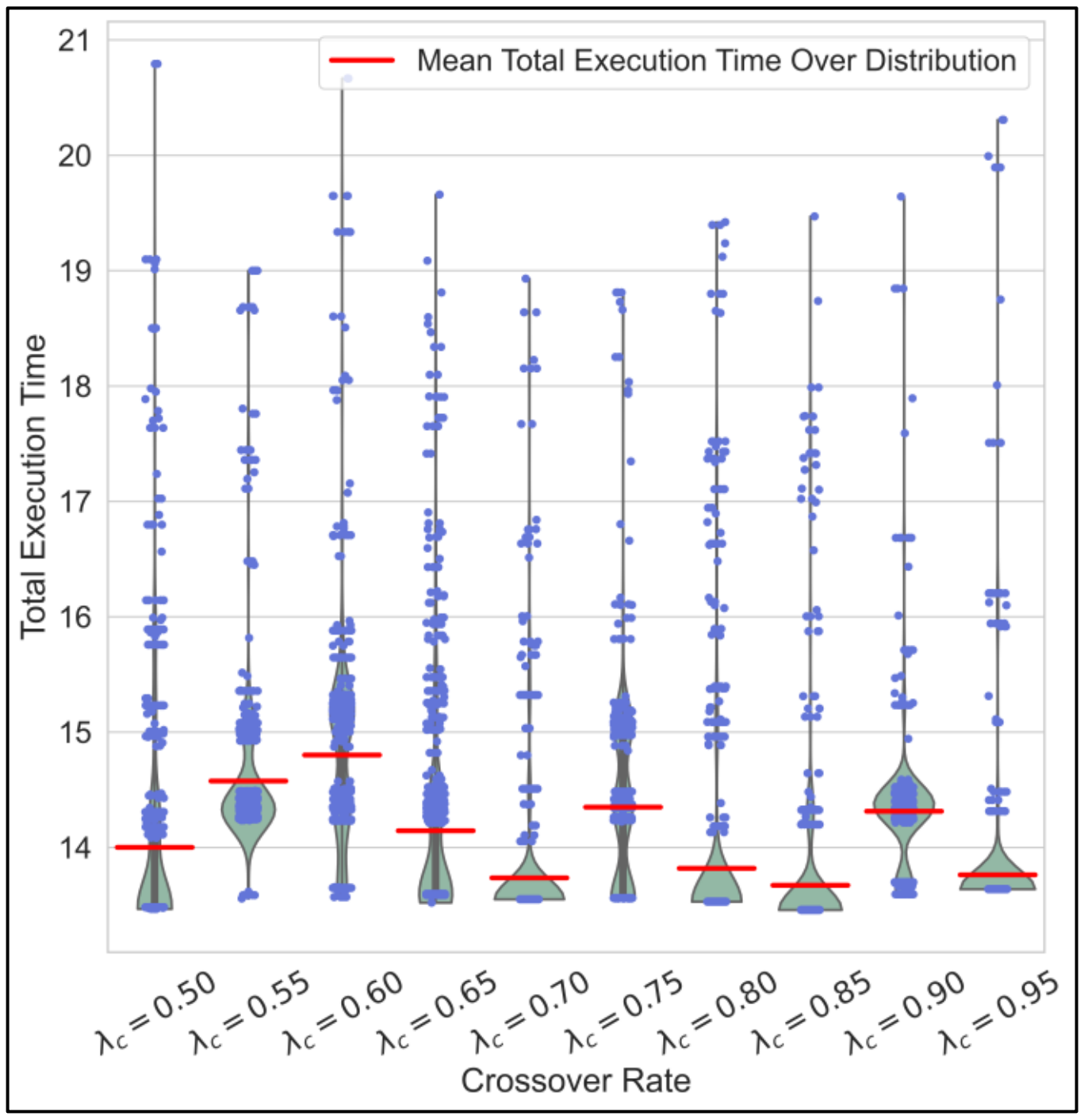

**Fig. 6.** Total execution time distribution over iterations for the requests' offloading solution using QoS-SLA-AGA versus crossover rate $\lambda_c$.

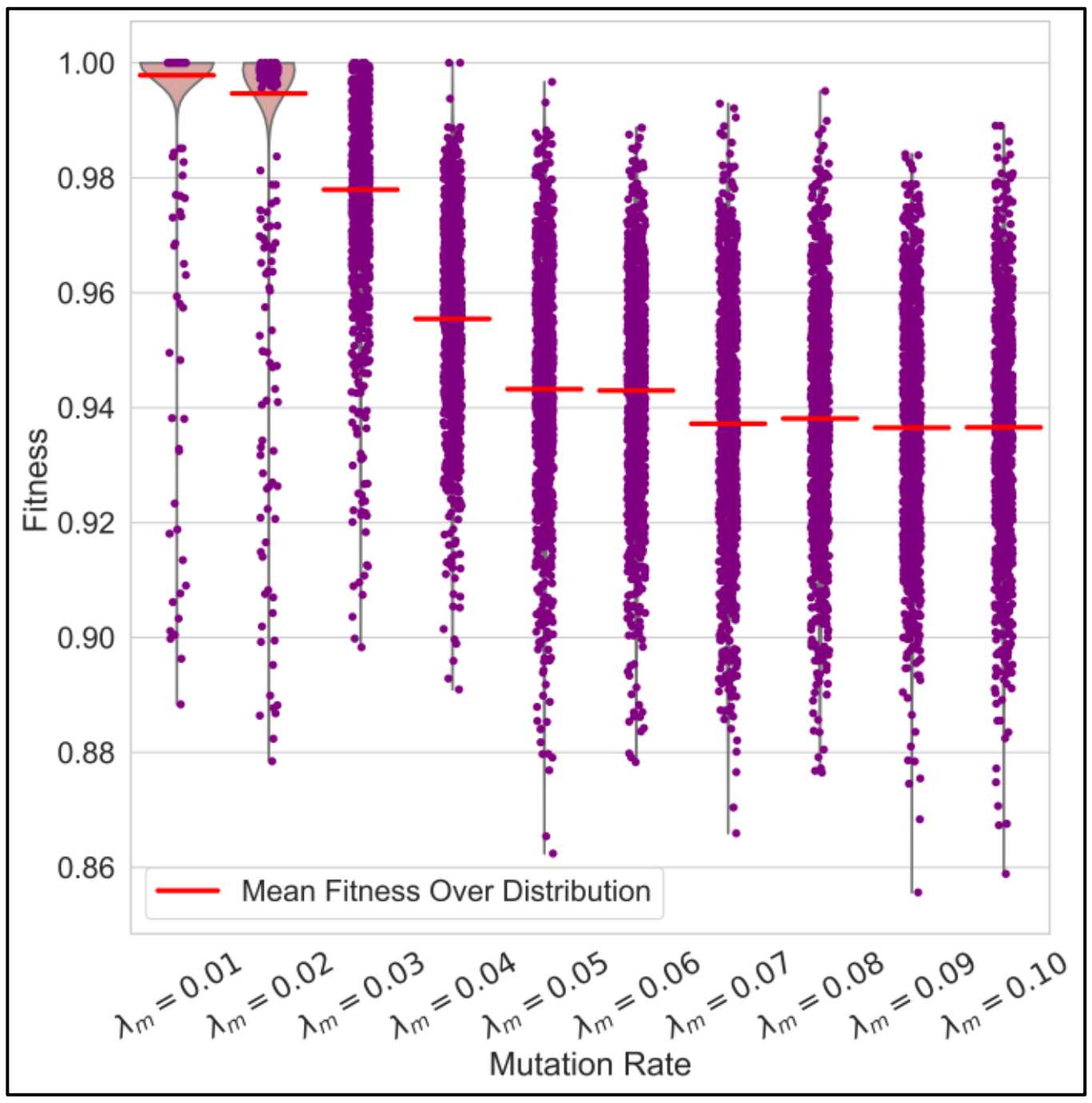

**Fig. 7.** Fitness score distribution over iterations for the requests' offloading solution using QoS-SLA-AGA versus mutation rate $\lambda_m$.

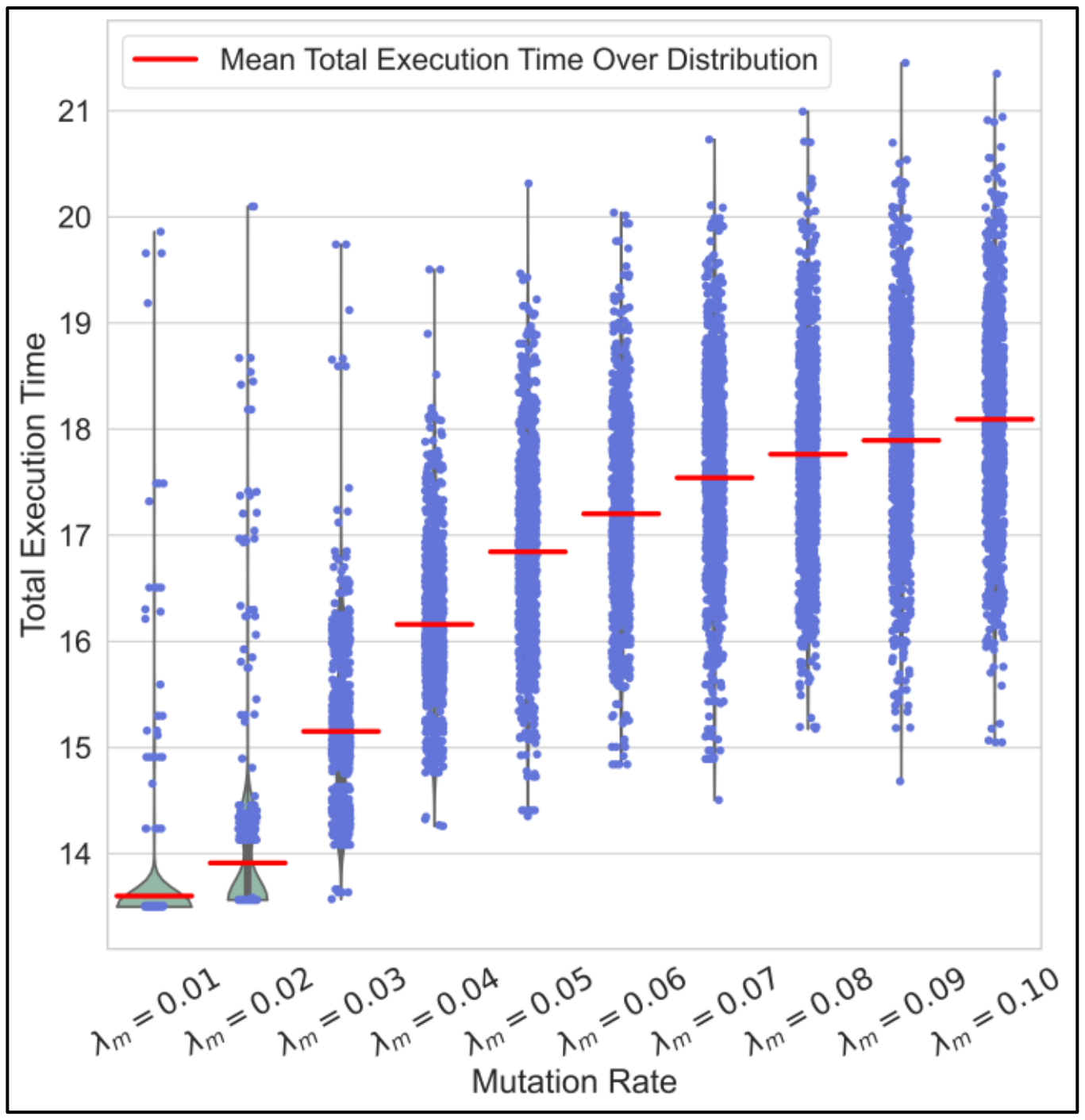

**Fig. 8.** Total execution time distribution over iterations for the requests' offloading solution using QoS-SLA-AGA versus mutation rate $\lambda_m$.

**Table V:** Optimal Values of Genetic Algorithm Parameters.

| Parameter | Value |
|---|---|
| Crossover rate ($\lambda_c$) | 0.95 |
| Mutation rate ($\lambda_m$) | 0.01 |
| Population size ($P_{size}$) | $2 \times requests$ |

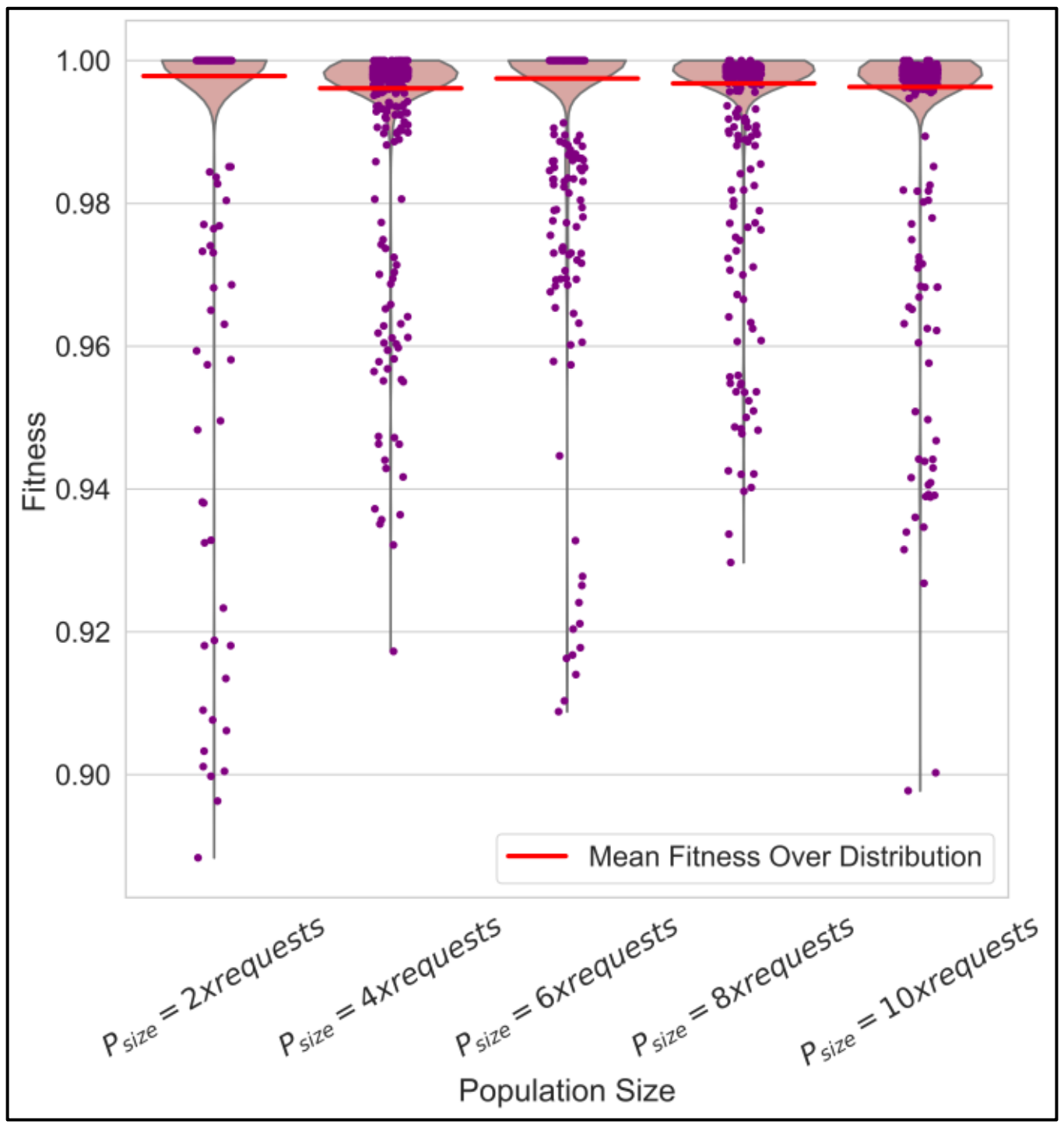

**Fig. 9.** Fitness score distribution over iterations for the requests' offloading solution using QoS-SLA-AGA versus population size $P_{size}$.

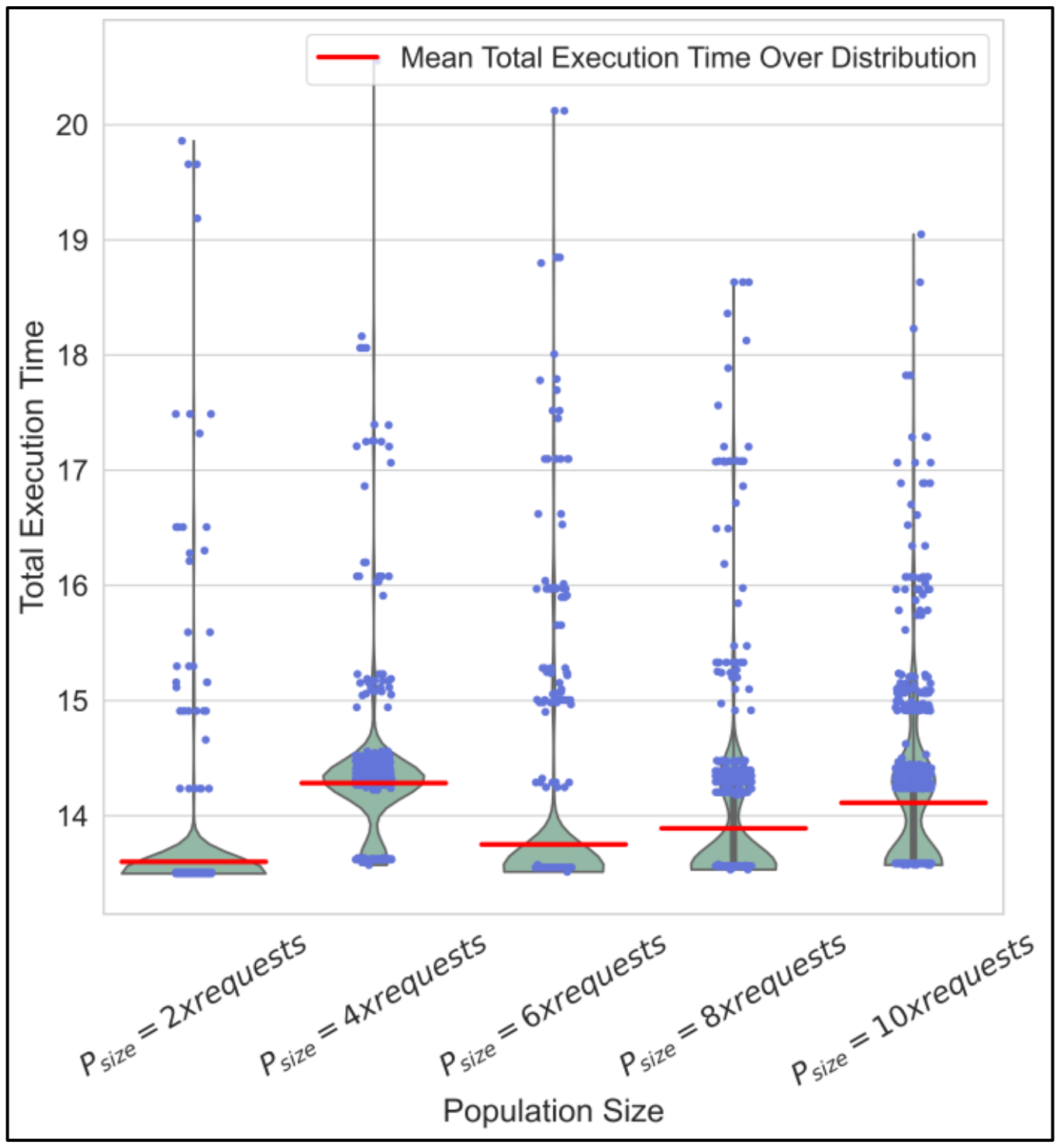

**Fig. 10.** Total execution time distribution over iterations for the requests' offloading solution using QoS-SLA-AGA versus population size $P_{size}$.

*C.2. Comparative Performance Analysis*

Figs. 11 and 12 show the total execution time and the number of requests violating SLA requirements respectively, for QoS-SLA-AGA, QoS-GA, and random offloading solutions with increasing latency requirements. As shown in Fig. 11 the random offloading algorithm has the longest execution time. While QoS-GA has better performance than the proposed QoS-SLA-AGA (Fig. 11), it has SLA violations (Fig. 12) compared to the proposed algorithm with no violations. This is because QoS-GA minimizes the total execution time without considering the SLA constraints. However, QoS-SLA-AGA minimizes the total execution time while considering the constraints. Thanks to its GA-based minimization strategy, the QoS-GA has fewer violations than the random offloading algorithm. In summary, the average total execution times with increasing latency requirements for QoS-SLA-AGA, QoS-GA, and random algorithms are 13.98 seconds, 13.52 seconds, and 19.58 seconds respectively. The average number of requests violating SLA constraints using QoS-SLA-AGA, QoS-GA, and random algorithms are 0, 1, 10.33 respectively. On average, the percentage of requests violating SLA constraints, with increasing latency requirements are 0, 5, and 51.66 using QoS-SLA-AGA, QoS-GA, and random approaches respectively.

Figs. 13 and 14 show the total execution time and the number of requests violating SLAs, respectively, for QoS-SLA-AGA, QoS-GA, and random offloading algorithms with increasing processing time requirements. As shown in Fig. 13 the random offloading algorithm has the longest execution time with increasing processing time requirements. Compared to QoS-GA, QoS-SLA-AGA results in a higher total execution time. This is because of the SLA constraints consideration in the proposed algorithm in addition to the objective of minimizing the total execution time. As shown in Fig. 14, QoS-SLA-AGA has no requests with SLA violations. However, the memory requirement is violated using QoS-GA, and the processing time, deadline, and memory requirements are violated using the random algorithm. As shown in the figure, the number of requests violating processing time requirements for the random approach decreases with increasing processing requirements. This is because more requests are executed within the processing time constraint. As a result, the number of requests violating the deadline requirement also decreases as the deadline requirement increases along with the processing time requirement. In summary, the average total execution times with increasing processing time requirements for QoS-SLA-AGA, QoS-GA, and random algorithms are 16.33 seconds, 13.50 seconds, and 19.21 seconds respectively. The average number of requests violating SLA constraints using QoS-SLA-AGA, QoS-GA, and random algorithms are 0, 1.33, 3.83 respectively. On average, the percentage of requests violating SLA constraints, with increasing processing requirements are 0, 6.66, and 19.16 using QoS-SLA-AGA, QoS-GA, and random approaches respectively.

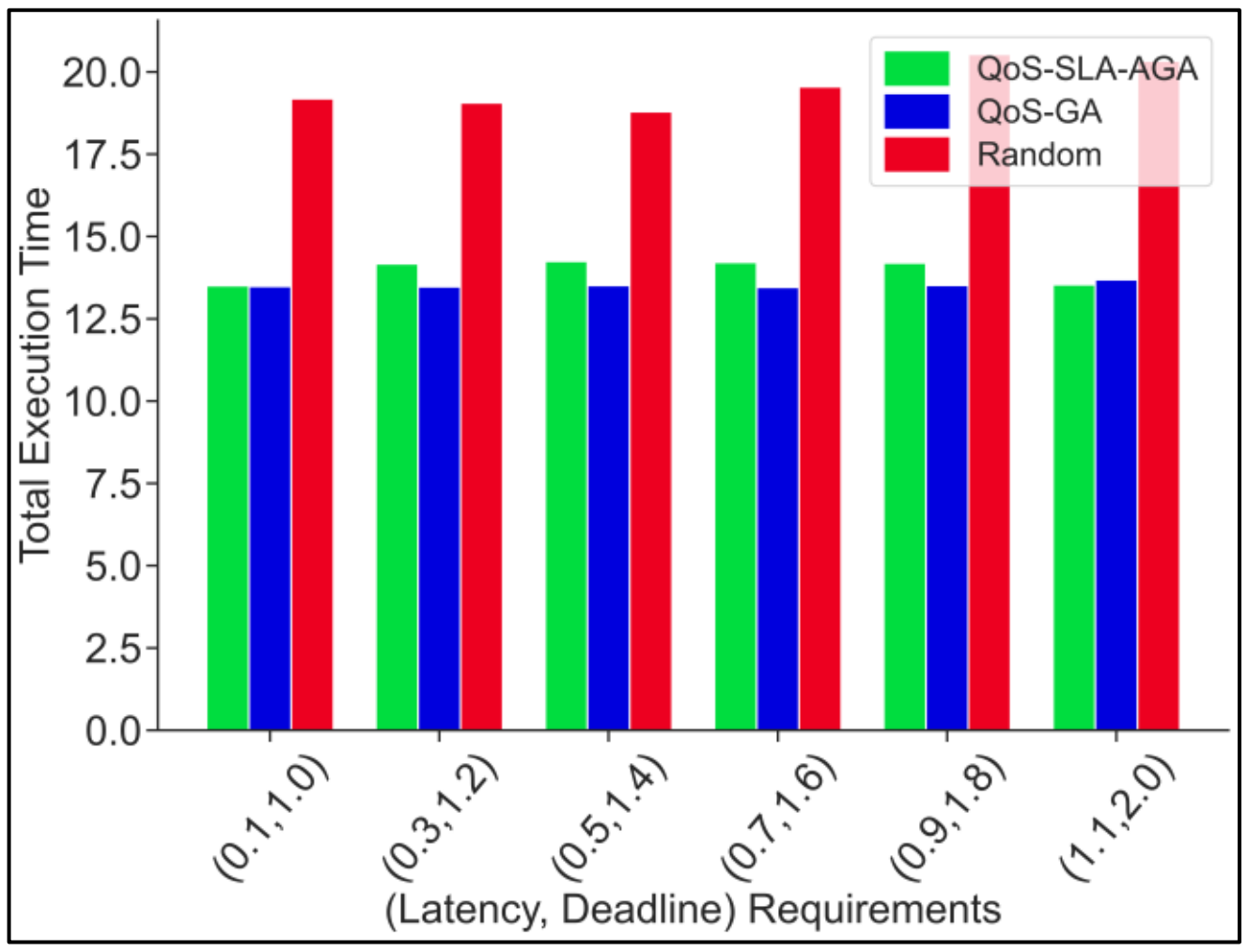

**Fig. 11.**Total execution time using QoS-SLA-AGA, QoS-GA, and random offloading algorithms versus latency requirements.

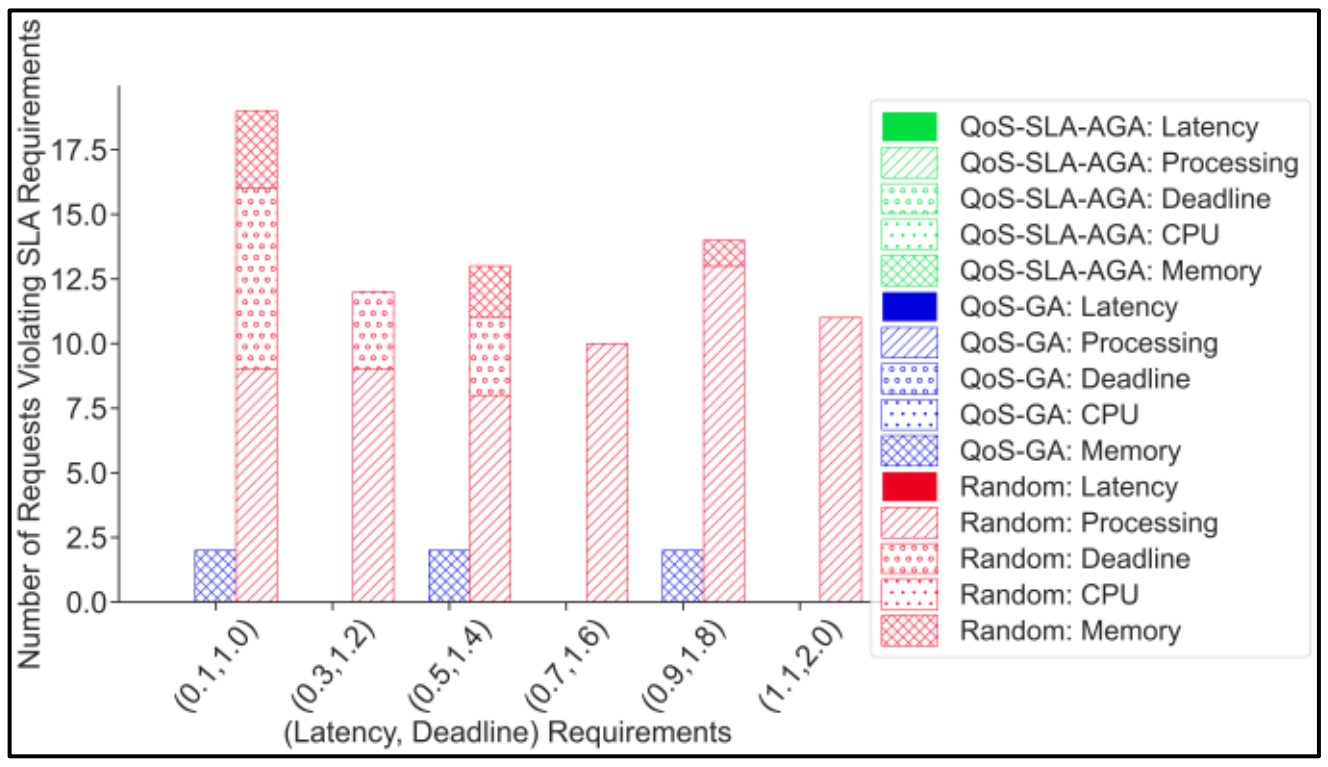

**Fig. 12.** Number of requests violating SLA constraints using QoS-SLA-AGA, QoS-GA, and random offloading algorithms versus latency requirements.

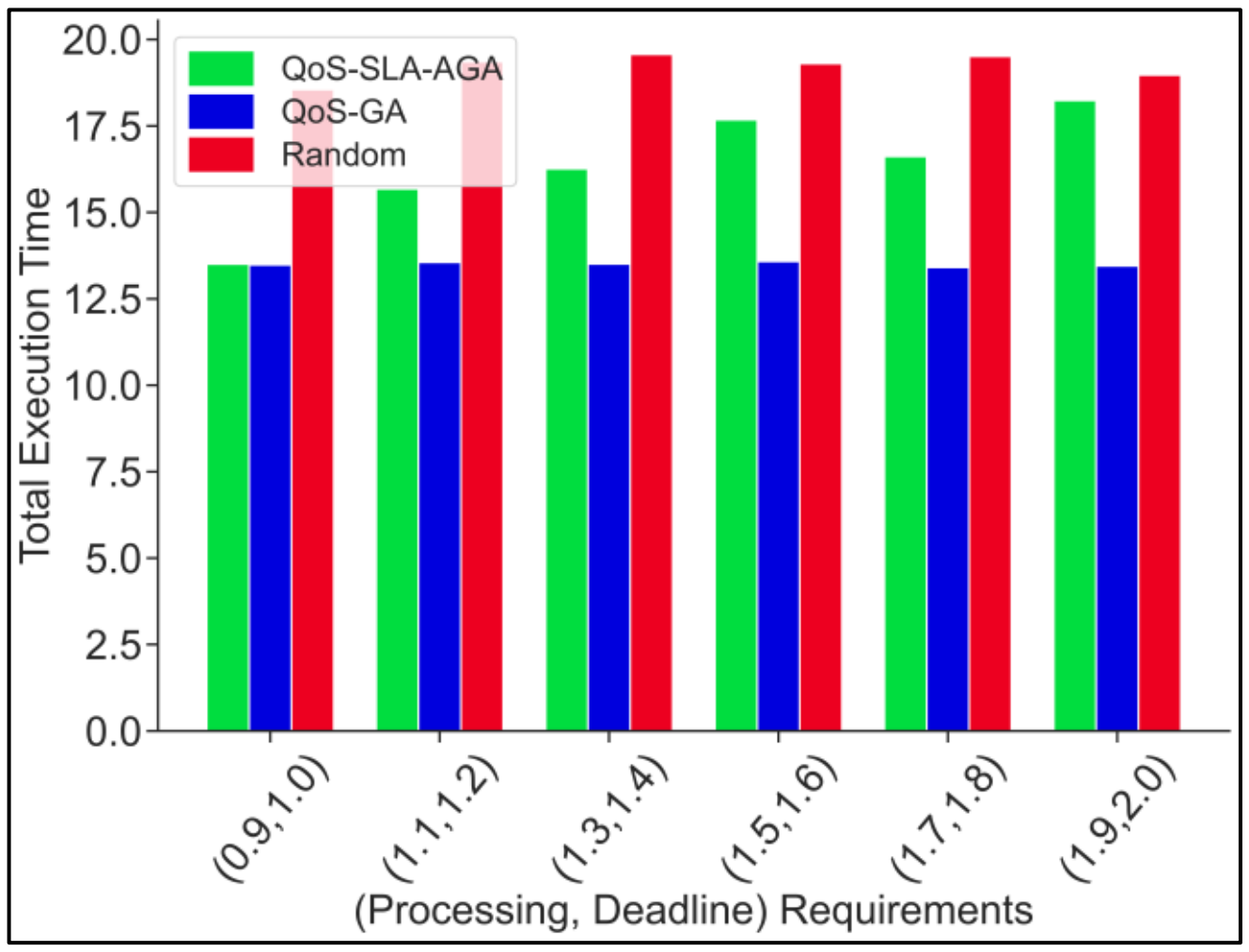

**Fig. 13.** Total execution time using QoS-SLA-AGA, QoS-GA, and random offloading algorithms versus processing time requirements.

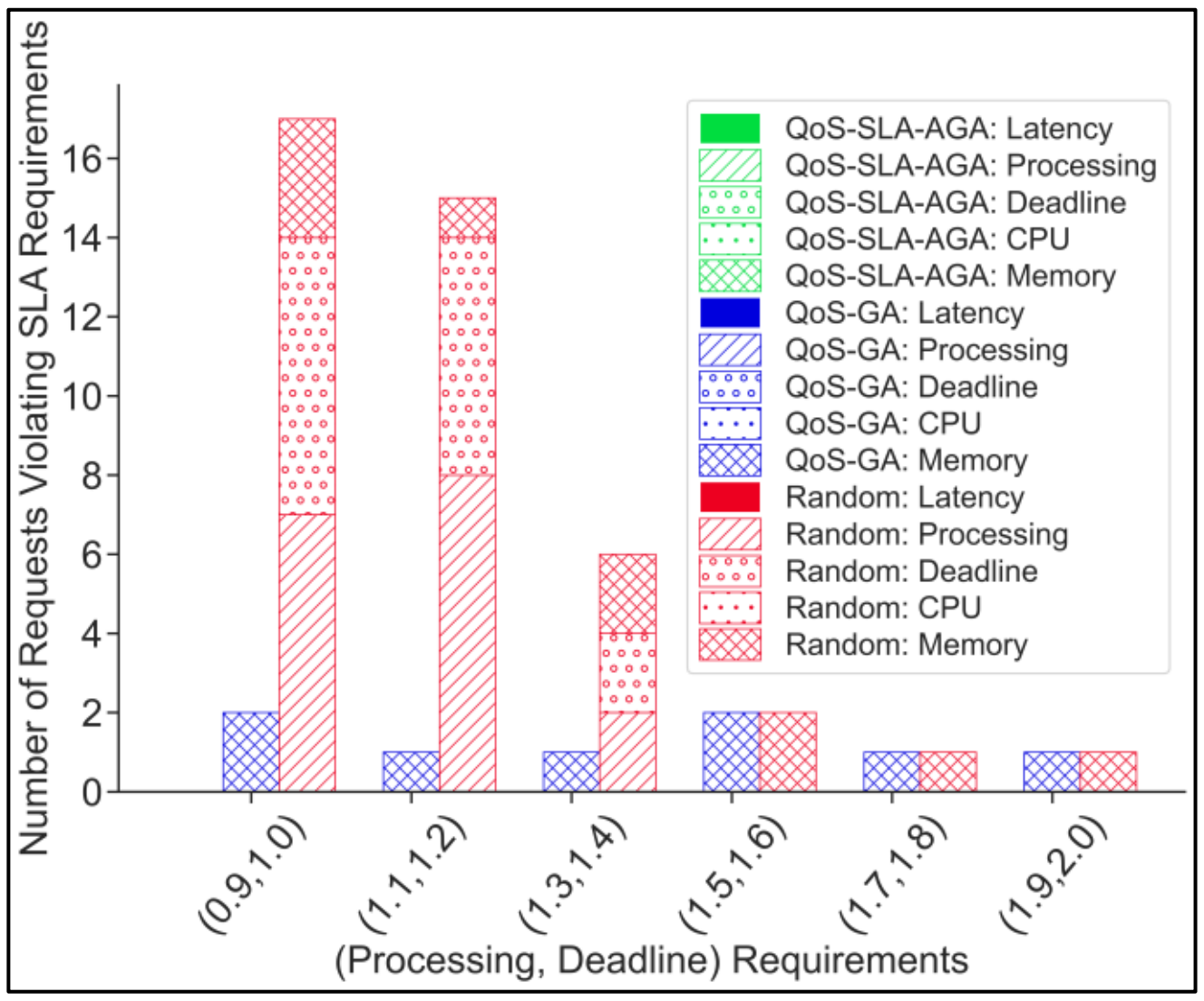

**Fig. 14.** Number of requests violating SLA constraints using QoS-SLA-AGA, QoS-GA, and random offloading algorithms versus processing time requirements.

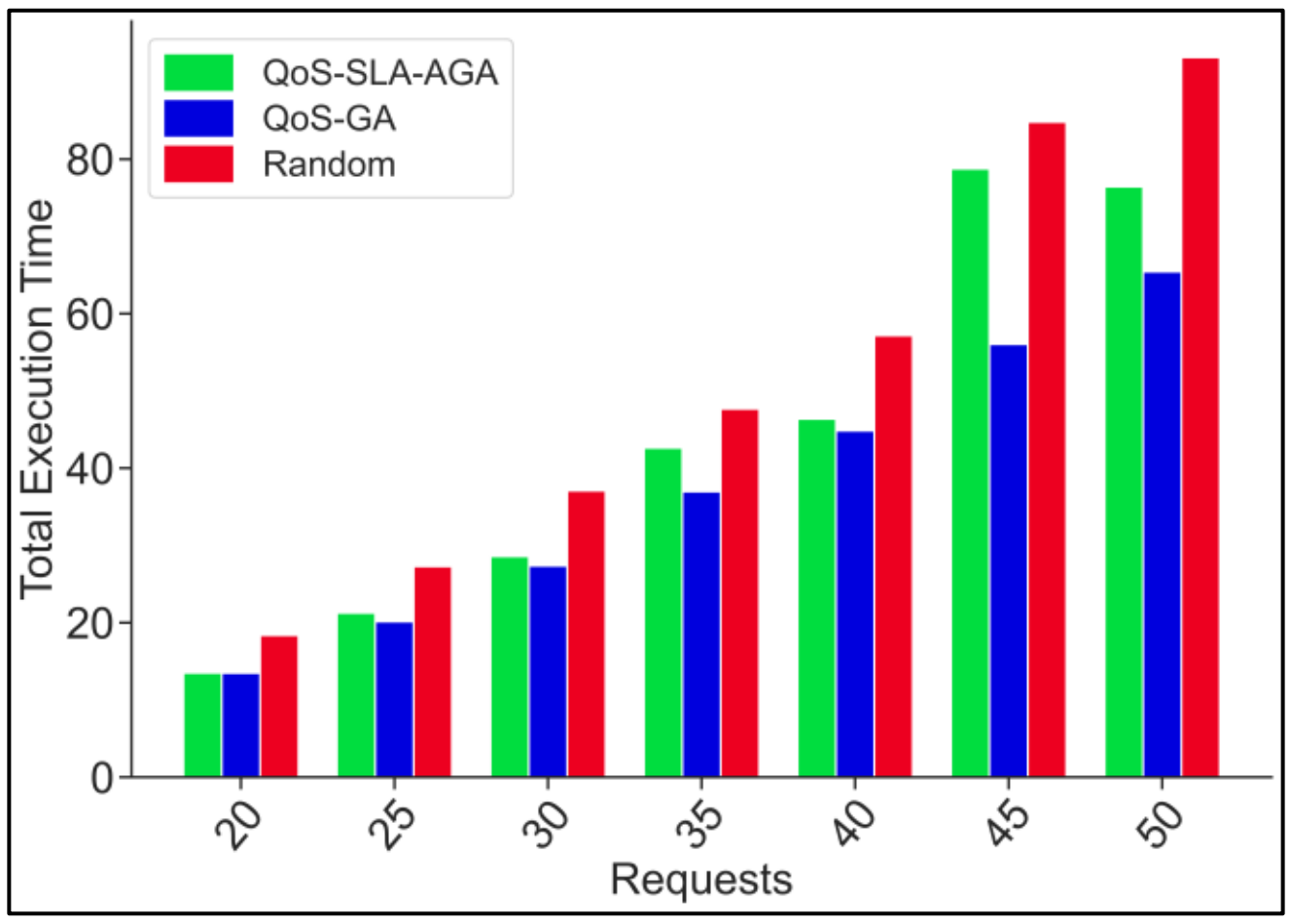

**Fig. 15.** Total execution time using QoS-SLA-AGA, QoS-GA, and random offloading algorithms versus the number of requests.

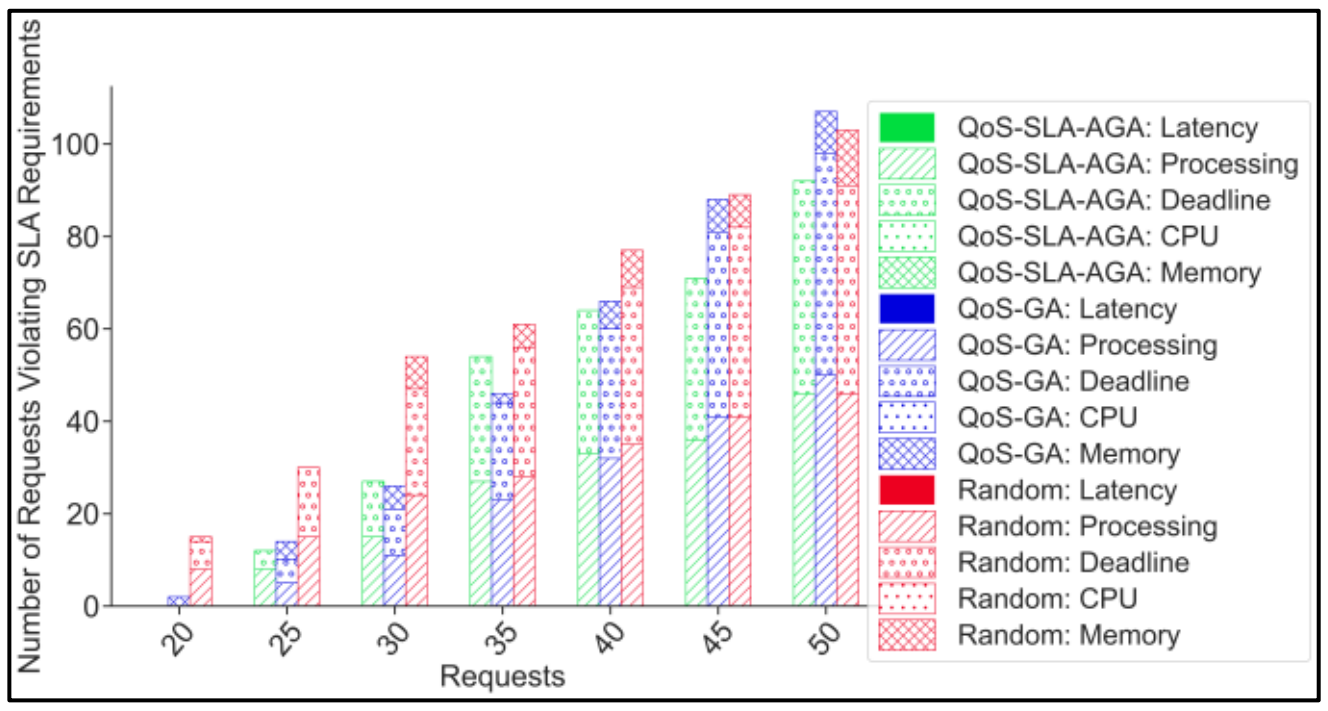

**Fig. 16.** Number of requests violating SLA constraints using QoS-SLA-AGA, QoS-GA, and random offloading algorithms versus the number of requests.

Figs. 15 and 16 show the total execution time and the number of requests violating SLAs, respectively, for QoS-SLA-AGA, QoS-GA, and random offloading algorithms with an increasing number of requests. As shown in Fig. 15 the random offloading algorithm has the longest execution time with an increasing number of requests, followed by QoS-SLA-AGA and QoS-GA algorithms. The total execution time of the QoS-GA algorithm is the minimum as it only focuses on minimizing the execution time without considering the SLA requirements of the requests. As shown in Fig. 16 all three algorithms violate the SLA requirements. The proposed algorithm violates the processing time and deadline requirements with an increasing number of requests. This is because the processing power of the servers is divided among requests with increasing requests while keeping the number of servers constant. Consequently, the processing time of requests increases leading to increased total execution time. However, in addition to processing and deadline requirements, the QoS-GA and the random algorithms violate the memory requirements of the requests with increasing requests. In summary, the average total execution times with an increasing number of requests using QoS-SLA-AGA, QoS-GA, and random algorithms are 43.93 seconds, 37.75 seconds, and 52.22 seconds respectively. QoS-SLA-AGA violates processing time and deadline requirements, whereas the QoS-GA and random algorithms violate the processing time, deadline, and memory requirements. In particular, the average number of requests violating SLA requirements using QoS-SLA-AGA, QoS-GA, and random algorithms are 23.71, 24, 28.60 respectively. On average, the percentage of requests violating SLA constraints with increasing requests are 59.40, 59.22, and 77.35 using QoS-SLA-AGA, QoS-AGA, and random approaches respectively.

## VII. Conclusion

Computation offloading is essential in an integrated edge-cloud system for IoV to enhance the QoS and respect SLA requirements of both compute-intensive and time-critical applications. In this paper, we propose QoS-SLA-AGA to offload vehicular applications' requests on an edge/cloud server such that the total execution time of the requests is minimized. Furthermore, our proposed optimization algorithm is constrained by the requests' SLA requirements in terms of latency, processing time, deadline, CPU, and memory. The proposed algorithm considers the overlapping of requests execution in the offloading decision. To the best of our knowledge, we are the first to propose a QoS-SLA-aware offloading algorithm using AGA in IoV that considers the overlapping of multi-request execution and dynamic speed of the vehicle for execution time minimization while adhering to the performance and resource SLA constraints. Numerical experiments and comparative analysis revealed that the proposed algorithm outperforms the random offloading approach in total execution time. In the context of SLA constraints, the proposed algorithm outperforms both baseline genetic-based and random offloading approaches. In future research, we propose to investigate QoS-SLA-aware partial offloading solutions where the request can be divided for simultaneous execution in an integrated vehicle-edge-cloud computing system.